\documentclass[twocolumn]{autart}    
\usepackage{cite}
\usepackage[pdftex]{graphicx}
\usepackage{amsmath,amssymb,amsfonts}
\usepackage{mathrsfs,mathtools}
\usepackage{dsfont}
\usepackage{array}
\usepackage{url}
\usepackage{color}
\usepackage{arydshln}
\usepackage{xspace}
\usepackage{float}
\usepackage{subcaption}
\usepackage{scrextend}
\usepackage{blindtext}
\usepackage{tikz}
\usepackage{xspace}
\usepackage{enumitem}
\usepackage{euscript}
\usepackage[normalem]{ulem}
\usepackage{wrapfig}
\usepackage{picins}

\def\extendedversion{1} 
\newcommand{\extver}[2]{%
  \ifx\extendedversion\undefined%
	#1%
  \else%
    #2%
  \fi%
}

\usepackage{csquotes}
\definecolor{mblue}{rgb}{0,0.4470,0.7410}
\newenvironment{definition}{\begin{defn}}{\end{defn}}
\newenvironment{theorem}{\begin{thm}}{\end{thm}}
\newenvironment{lemma}{\begin{lem}}{\end{lem}}
\newenvironment{corollary}{\begin{cor}}{\end{cor}}
\newenvironment{example}{\begin{exmp}}{\end{exmp}}
\newenvironment{remark}{\begin{rem}}{\end{rem}}
\newenvironment{proof}{%
	\extver{\vspace{-3mm}}{}%
	\begin{pf}%
}{%
	\hfill$\blacksquare$
	\end{pf}
	\extver{\vspace{-2mm}}{}%
}

\newcommand{\norm}[1]{\left\lVert#1\right\rVert}

\newcommand{\ltwo}{\ensuremath{\mathcal{L}_2}\xspace}
\newcommand{\lone}{\ensuremath{\mathcal{L}_1}\xspace}

\newcommand{\litwo}{\ensuremath{\mathcal{L}_{\mathrm{i}2}}\xspace}
\newcommand{\linf}{\ensuremath{\mathcal{L}_\infty}\xspace}
\newcommand{\liinf}{\ensuremath{\mathcal{L}_{\mathrm{i}\infty}}\xspace}
\newcommand{\mc}[1]{\mathcal{#1}}
\newcommand{\mb}[1]{\mathbb{#1}}
\newcommand{\mr}[1]{\mathrm{#1}}
\newcommand{\ms}[1]{\mathscr{#1}}
\newcommand{\mf}[1]{\mathfrak{#1}}

\newcommand{\htwo}{\ensuremath{\mathcal{H}_2}\xspace}
\newcommand{\hitwo}{\ensuremath{\mathcal{H}^{\mathrm{g}}_{\mathrm{i}2}}\xspace}

\newcommand{\dif}{\mathrm{d}}
\newcommand{\differ}[2]{\frac{\dif #1}{\dif #2}}

\newcommand{\Partial}[3]{\frac{\partial^{#1} #2}{\partial #3^{#1}}}
\newcommand{\tPartial}[3]{\tfrac{\partial^{#1} #2}{\partial #3^{#1}}}
\newcommand{\Differ}[3]{\frac{\dif^{#1} #2}{\dif #3^{#1}}}

\newcommand{\coldef}[1]{\ensuremath{\mathrm{col}\!\left(#1\right)}}

\DeclareMathOperator*{\arginf}{arg\,inf}

\newcommand{\diffstate}[2]{\begin{pmatrix}#1-\tilde{#1}\\#2-\tilde{#2}\end{pmatrix}}

\newcommand{\shrtranspose}{^{\!\top\!}}
\newcommand{\posdef}{\succ}
\newcommand{\negdef}{\prec}
\newcommand{\possemidef}{\succcurlyeq}
\newcommand{\negsemidef}{\preccurlyeq}
\newcommand{\unaryminus}{\scalebox{0.65}[1.0]{\ensuremath{\,-\,}}}
\newcommand{\spacenafig}{\extver{\vspace{-6mm}}{}}
\newcommand{\spacenacap}{\vspace{-0mm}}
\newcommand{\setdefinition}[2]{\ensuremath{\left\{ \vphantom{#2} #1 \right. \, \left| \, \vphantom{#1} #2 \right\}}}

\renewcommand{\det}[1]{\mathrm{det}\!\left(#1\right)}

\newcommand{\supply}{\EuScript{S}}
\newcommand{\storage}{\EuScript{V}}
\newcommand{\QSR}{\begin{pmatrix} Q  & S\\ S^\top & R \end{pmatrix}}
\newcommand{\cohull}[1]{\mr{co}\{#1\}}

\newcounter{ass}
\newcommand{\assume}[2]{\refstepcounter{ass}\begin{itemize}
\item[A\theass] #1 \label{A:#2}
\end{itemize}}

\newcommand{\unitstep}{\mathds{1}(t)}

\newcommand{\integral}[2]{\extver{{\textstyle \int_{#1}^{#2}\,}}{\int_{#1}^{#2}}}

\newcommand{\biographytwo}[3]{
\par\noindent 
\parbox[t]{\linewidth}{
\noindent\parpic{\includegraphics[width=.85in,keepaspectratio]{#1}}
\noindent {\scriptsize  {\bf #2}\
#3}}
}

\newcommand{\comment}[1]{}

\newcommand{\added}[1]{#1}

\extver{
\setlength{\parskip}{.3em}
\let\OLDthebibliography\thebibliography
\renewcommand\thebibliography[1]{
  \OLDthebibliography{#1}
  \setlength{\parskip}{0pt}
  \setlength{\itemsep}{0pt plus 0.1ex}
}
\captionsetup{font=small}
\captionsetup[sub]{font=footnotesize}
\makeatletter
\g@addto@macro\normalsize{%
  \setlength\abovedisplayskip{3pt plus 1pt minus 1pt}
  \setlength\belowdisplayskip{\abovedisplayskip}
  \setlength\abovedisplayshortskip{0pt plus 1pt}
  \setlength\belowdisplayshortskip{\abovedisplayskip}
}

\newcommand{\shortsection}[1]{\vspace{-1mm}\section{#1}\vspace{-1mm}}
\newcommand{\shortsubsection}[1]{\vspace{-.8mm}\subsection{#1}\vspace{-1.2mm}}
}{
\newcommand{\shortsection}[1]{\section{#1}}
\newcommand{\shortsubsection}[1]{\subsection{#1}}
}
\newcommand{\sche}{p}

\begin{document}

\begin{frontmatter}
\title{Convex Incremental Dissipativity Analysis \mbox{of Nonlinear Systems\thanksref{footnoteinfo}}}
\extver{\vspace{-5mm}}{\subtitle{\small (Extended version)}}

\thanks[footnoteinfo]{This paper was not presented at any IFAC meeting. This work has received funding from the European Research Council (ERC) under the European Union's Horizon 2020 research and innovation programme (grant agreement nr. 714663) {and was also supported by the European Union within the framework of the National Laboratory for Autonomous Systems (RRF-2.3.1-21-2022-00002).} 
Corresponding author: C.~Verhoek.
\emph{Email addresses:} \textbraceleft \texttt{c.verhoek}, \texttt{p.j.w.koelewijn}, \texttt{s.haesaert}, \texttt{r.toth}\textbraceright \texttt{@tue.nl}.%
}


\author[tue]{Chris Verhoek}, 
\author[tue]{Patrick J. W. Koelewijn}, 
\author[tue]{Sofie Haesaert}, 
\author[tue,sztaki]{Roland T\'oth} 

\extver{\vspace{-1mm}}{}

\address[tue]{Control Systems Group, Eindhoven University of Technology, Eindhoven 5600MB, The Netherlands}
\extver{\vspace{-1mm}}{}
\address[sztaki]{Systems and Control Lab, Institute for Computer Science and Control, 1111 Budapest, Hungary}
\extver{\vspace{-2mm}}{}
\begin{abstract}
\vskip -4mm 
\noindent Efficiently computable stability and performance analysis of nonlinear systems becomes increasingly more important in practical applications. Dissipativity can express stability and performance jointly, but existing results are limited to the regions around the equilibrium points of these nonlinear systems. The incremental framework, based on the convergence of the system trajectories, removes this limitation. We investigate how stability and performance characterizations of nonlinear systems in the incremental framework are linked to dissipativity, and how general performance characterization beyond the \ltwo-gain concept can be understood in this framework. This paper presents a matrix inequalities-based convex incremental dissipativity analysis for nonlinear systems via quadratic storage and supply functions. The proposed dissipativity analysis links the notions of incremental, differential, and general dissipativity. We show that through differential dissipativity, incremental and general dissipativity of the nonlinear system can be guaranteed. These results also lead to the incremental extensions of the \ltwo-gain, the generalized \htwo-norm, the \linf-gain, and passivity of nonlinear systems.
\vskip -4mm
%
%
\end{abstract}
\end{frontmatter}
%

\shortsection{Introduction}
The \emph{linear time-invariant} (LTI) framework has been a systematic and easy-to-use {approach} for modeling, identification and control of physical systems for many years. Its success is driven by powerful theoretical and computational  results on stability, performance, and shaping \cite{Skogestad}. Growing performance demands in terms of accuracy, response speed and energy efficiency, together with increasing complexity of systems to accommodate such expectations, are pushing beyond the modeling and control {capabilities} of the LTI framework. Therefore, stability and performance analysis of \emph{nonlinear} systems becomes increasingly more important.

A large variety of stability analysis tools are available for nonlinear systems, including Lyapunov's stability theory \cite{KhalilNL}, dissipativity theory \cite{Willems1972} and contraction theory \cite{Lohmiller1998}. Moreover, techniques such as backstepping, input-output or feedback linearization  \cite{KhalilNL} have been {introduced} to stabilize the behavior and to achieve reference tracking for nonlinear systems. However, these techniques often require cumbersome computations and restrictive assumptions, and -- unlike the LTI case -- they have not lead to systematic performance analysis and shaping methods. While dissipativity theory in principle allows for analysis of nonlinear systems, current results are not computationally attractive. Furthermore, they only provide local stability and performance guarantees, i.e., only w.r.t. a single point of natural storage (usually the origin), which is undesirable for disturbance rejection and reference tracking. Hence, there is need for a computationally efficient analysis tool for \emph{global} conclusions on the dissipativity property of a nonlinear system.

Several frameworks have been developed to extend computationally efficient LTI tools to nonlinear systems, e.g., using piece-wise affine, \emph{linear time-varying} (LTV), Fuzzy, or \emph{linear parameter-varying} (LPV) system representations. The LPV framework specifically aims at providing convex tools to analyze nonlinear systems as a predefined convex set of LTI systems. However, the stability and performance guarantees are still only valid w.r.t. a single equilibrium point \cite{KoelewijnMazzoTothWeiland2019}. To analyze global stability properties of nonlinear systems, independent of a specific equilibrium point, notions such as incremental stability \cite{angeli2002lyapunov} were introduced. Incremental stability analyzes stability of a system w.r.t. arbitrary trajectories of the system, instead of w.r.t. a single equilibrium point. \added{Similar stability notions have also been developed, such as contraction \cite{Lohmiller1998, Manchester2018} and convergence theory \cite{PavlovWouwNijmeijer2004} with strong connections to incremental stability theory \cite{Ruffer2013}. Similar notions for performance have also been introduced such as incremental \ltwo-gain \cite{FromionMonacoCyrot2001} and passivity \cite{PavlovMarconi2007}.}  Extensions towards global dissipativity analysis in {the} literature are differential dissipativity \cite{vanderSchaft2013, ForniSepulchre2013, ForniSepulchreSchaft2013}, incremental dissipativity \cite{PavlovMarconi2007} and equilibrium independent dissipativity \cite{SimpsonPorco}. However, they do not provide computationally efficient methods to verify these dissipativity notions. Works discussing differential and incremental dissipativity only focus on passivity-based performance and how the various dissipativity notions are linked to general dissipativity is generally not discussed.

To address these shortcomings, the main contributions of this paper are (i) conditions on general quadratic performance analysis using incremental dissipativity, (ii) establishing the missing link between general dissipation theory and incremental analysis {of} nonlinear systems, and (iii) computationally efficient convex tools to analyze incremental stability and performance of nonlinear systems. This is achieved by developing a general incremental dissipativity framework that connects differential dissipativity, incremental dissipativity and general dissipativity. As a consequence, incremental notions of the \ltwo-gain, the generalized \htwo-norm, the \linf-gain and passivity are systematically introduced also recovering of some  existing results on {these} concepts. Furthermore, convex analysis tools to compute the resulting conditions for differential and incremental dissipativity are derived using a so-called \emph{differential parameter-varying} (DPV) inclusion of the nonlinear system.

In Section~\ref{section:problemdef}, a formal definition of the problem setting is given. Section~\ref{section:mainresults} gives the main results on differential, incremental and general dissipativity and their connection. In Section~\ref{section:applications}, the incremental extensions of well-known performance measures are derived and the concept of {DPV} inclusions are discussed, yielding convex computation methods. The introduced concepts and methods are demonstrated on \extver{an}{two} academic example\extver{}{s} in Section~\ref{section:example}, while the conclusions are provided Section~\ref{section:conclusions}.

\begin{footnotesize}
\extver{\vspace{-3mm}}{}\subsubsection*{Notation.}\extver{\vspace{-1mm}}{}
$\mathbb{R}$ is the set of real numbers, while $\mathbb{R}_0^+$ and $\mathbb{R}^+ $ stand for non-negative reals and positive reals. 
The convex hull of a set $\mc{S}$ is $\cohull{\mc{S}}$. 
Projection of $\mb{D}:=\mb{A}\times\mb{B}$, with elements $(a,b)$, onto $\mb{A}$ is denoted by $\pi_a\mb{D}$, {meaning} $a\in\pi_a\mb{D}=\mb{A}$. 
If a mapping $f:\mathbb{R}^p\to\mathbb{R}^q$ is in $\mc{C}^n$, it is $n$-times continuously differentiable. 
$\mathscr{L}^n_2$ {is} the signal space {of} real-valued square integrable functions $f:\mathbb{R}_0^+\to\mathbb{R}^n$ with associated norm $\norm{f}_2:= (\int_0^{\infty}\|f(t)\|^2\hspace{0.4mm} \dif t)^{\frac{1}{2}}$ where $\|\cdot\|$ is the Euclidean (vector) norm. $\mathscr{L}^n_\infty$ {is} the signal space of functions $f:\mathbb{R}^+_0\to\mathbb{R}^n$ with finite amplitude, i.e. bounded $\norm{f}_\infty\!\! := \sup_{t\ge0}\norm{f(t)}$. 
We use $(*)$ to denote a symmetric term {in a quadratic expression, e.g. $(*)\shrtranspose Q(a-b) = (a-b)\shrtranspose Q(a-b)$ for $Q\in\mathbb{R}^{n\times n}$ and $a,b\in\mathbb{R}^{n}$}. 
The notation $A\posdef0$ ($A\possemidef0$) indicates that $A$ is positive (semi-) definite, while $A\negdef0$ ($A\negsemidef0$) {denotes a} negative (semi-) definite $A$. 
The zero-matrix and the identity matrix of appropriate dimensions are denoted as $0$ and $I$. 
Furthermore, $\coldef{x_1, \dots ,x_n}$ denotes the column vector $[x_1^\top \cdots x_n^\top]^\top$. 
\end{footnotesize}

\shortsection{Problem definition}\label{section:problemdef}
In this paper, we consider nonlinear, time-invariant systems of the form
\begin{equation}\label{eq:sys}
	\Sigma : \left\{
	\begin{aligned}
		\dot{x}(t)&=f(x(t),u(t));\\ y(t) &= h(x(t),u(t));
	\end{aligned}
	\right.
\end{equation}
where $x(t)\in \mc{X} \subseteq \mb{R}^{n_\mr{x}}$ is the state, $u(t)\in\mc{U} \subseteq \mb{R}^{n_\mr{u}}$ is the input, and $y(t)\in\mc{Y} \subseteq \mb{R}^{n_\mr{y}}$ is the output 
of the system. The sets $\mc{X}$, $\mc{U}$ and $\mc{Y}$ are open sets containing the origin, {with  $\mc{X}$, $\mc{U}$ being convex,} and the mappings $f: \mc{X} \times \mc{U}\to \mb{R}^{n_\mr{x}}$ and $h: \mc{X} \times \mc{U}\to \mc{Y}$ are in $\mc{C}^1$. 
{We} only consider solutions of \eqref{eq:sys} that are forward complete, unique and satisfy \eqref{eq:sys} in the ordinary sense. The trajectories of \eqref{eq:sys} are also restricted to have left-compact support, i.e., $\exists\, t_* \in \mb{R}$ such that {$(x,u,y)$} is zero outside the left-compact set $[t_*,\infty)$. We define the state-transition map as $\phi_{\mr{x}}:\mb{R}\times\mb{R}\times\mc{X}\times\mc{U}^\mb{R}\to\mc{X}$, {describing} the evolution of the state such that
\begin{equation}
x(t) = \phi_{\mr{x}}(t, t_0, {x_0}, u),
\end{equation}
with $x_0 = x(t_0)$.
The behavior of the system, i.e., the set of all possible solutions, is denoted by
\begin{multline}
\hspace{-1mm} \mathfrak{B} := \{(x,u,y)\in\left(\mc{X}\times\mc{U}\times\mc{Y}\right)^{\mb{R}}\mid x\in\mc{C}^1\, \text{and} \left(x,u,y\right) \\ 
\text{satisfies \eqref{eq:sys} with left-compact support}\}.
\end{multline}
Note that $\mathfrak{B}\subseteq \mathcal{B}^\mathbb{R}$, where ${\mathcal{B}}
= \mc{X}\times\mc{U}\times\mc{Y}$ is called the signal value set.  

In this paper, the form presented in \eqref{eq:sys} will be referred to as the \emph{primal form} of the nonlinear system.
For {the primal form,} an extensive dissipativity theory has been developed over the years, with its roots in \cite{Willems1972}. From the notion of dissipativity, many system properties can be derived, such as performance characteristics and stability \cite{Willems1972, hill1980dissipative}, as well as a link with the physical interpretation of the system. Therefore, dissipativity is an important fundament in nonlinear system theory, {which we will briefly review}. We consider Willems' dissipativity notion 
\cite{Willems1972} {that} allows for simultaneous stability and performance analysis.
\begin{definition}[General dissipativity]
The system \eqref{eq:sys} is dissipative w.r.t. a \emph{supply} function $\supply: \mc{U} \times \mc{Y} \to \mb{R}$, if there exists a \emph{storage} function $\storage: \mc{X} \to \mb{R}_0^+$ with an $x_\ast\in\mc{X}$, such that $\storage(x_\ast) = 0$ and
\begin{equation}
\storage(x(t_1)) - \storage(x(t_0)) \le \integral{t_0}{t_1}\supply(u(t),y(t))\dif t,
\label{eq:DIE}
\end{equation}
for all $t_0,t_1\in\mb{R}$ with $t_0 \le t_1$, {and for all $(x,u,y)\in\mf{B}$}. 
\end{definition}
The storage function $\storage$ can be interpreted as a representation of the stored `energy' in the system with a point of neutral storage $x_\ast$ (energy minimum), while the supply function $\supply$ can be seen as the total energy flowing in and out of the system. If $\storage(x(t))$ is differentiable, the \emph{dissipation inequality} {(DI)} \eqref{eq:DIE} can be rewritten as the so-called \emph{differentiated dissipation inequality} (DDI), i.e., $\differ{}{t}\big(\storage(x(t))\big) \le \supply(u(t),y(t))$. In this paper, dissipativity of the primal form of a system will be referred to as \emph{general dissipativity}.
Note that $x_\ast$, i.e., the point where $\storage$ is considered to be zero, does not need to be at $x_\ast=0$. In fact, it can be chosen to be any (forced) equilibrium point of \eqref{eq:sys}. However, if the system is nonlinear, the DDI is different for each considered $x_\ast$ and unlike in the LTI case, this difference cannot be eliminated by a coordinate transformation. This means that performance and stability analysis through general dissipativity is equilibrium point \emph{dependent}.

\begin{figure}[!t]
\centering
\includegraphics[width=\linewidth]{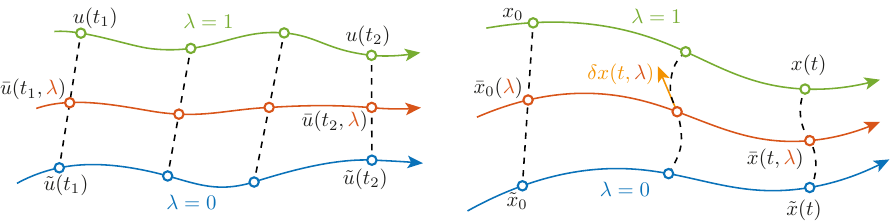}
\spacenafig
\caption{Collection of trajectories $(\bar{x}{(\lambda)},\bar{u}{(\lambda)},\bar{y}{(\lambda)})\in\mf{B}$ parametrized in  $\lambda\in[0,1]$ between two arbitrary state and input trajectories, where $\bar{u}(t,\lambda) = \tilde{u}(t)+\lambda(u(t)-\tilde{u}(t))$ and $\bar{x}_0(\lambda) = \tilde{x}_0+\lambda(x_0-\tilde{x}_0)$. Variation $\delta x$ of $\bar{x}$ is given in orange.}
\spacenacap
\label{fig:trajectories}
\end{figure}

An extension to this concept is \emph{incremental dissipativity}, i.e., analysis of the (dissipated) energy flow between any two system trajectories. We give an extension of the definition of incremental passivity in \cite[Def. 4.7.1]{L2PassiveSchaft2017}:
\begin{definition}[Incremental Dissipativity]\label{def:incrementaldissipativity} The system { \eqref{eq:sys}} is called incrementally dissipative w.r.t. the supply function $\supply_\Delta:\mc{U}\times\mc{U}\times\mc{Y}\times\mc{Y}\to\mb{R}$, if there exists a storage function ${\storage_\Delta}: \mc{X}\times\mc{X}\to\mb{R}_0^+$, with ${\storage_\Delta}(x,x)=0$, such that for any two trajectories $(x,u,y),(\tilde{x},\tilde{u},\tilde{y})\in\mathfrak{B}$,
\begin{multline}
{\storage_\Delta}\big(x(t_1), \tilde{x}(t_1)\big) - {\storage_\Delta}\big(x(t_0), \tilde{x}(t_0)\big) \\ \le \integral{t_0}{t_1}\supply_\Delta \big(u(t), \tilde u(t),y(t), \tilde y(t)\big)\dif t,
\label{eq:IDIE}
\end{multline}
for all $t_0,t_1\in\mb{R}$ with $t_0 \le t_1$. 
\end{definition}
Besides analyzing the difference between two trajectories, it is also possible to analyze \emph{infinitesimal} variations of trajectories. 
First, define \added{the set of paths}
\begin{equation*}
	\Gamma_\Phi(\varphi,\tilde \varphi) := \lbrace \bar{\varphi} \in \Phi^{[0,1]}\mid \bar \varphi\in\mathcal{C}^1,\, \bar{\varphi}(0) = \tilde \varphi,\, \bar{\varphi}(1)=\varphi\rbrace,
\end{equation*}
corresponding to all smooth paths along $\lambda\in[0,1]$ between points $\varphi\in \Phi\subseteq\mathbb{R}^\mr{n}$ and $\tilde \varphi\in \Phi\subseteq\mathbb{R}^\mr{n}$. Consider any two trajectories of \eqref{eq:sys}:
$(x,u,y),(\tilde{x},\tilde{u},\tilde{y})\in\mf{B}$. As $\mc{X}$, $\mc{U}$ are considered to be convex, any trajectory \emph{between} these can be parametrized with
$\bar{x}_0\in\Gamma_\mc{X}(x_0,\tilde x_0)$ and $\bar{u}(t)\in\Gamma_\mc{U}(u(t),\tilde u(t))$, e.g., $\bar{x}_0(\lambda) = \tilde{x}_0+\lambda(x_0-\tilde{x}_0) \added{\in\mc{X}}$ and $\bar{u}(t,\lambda) = \tilde{u}(t)+\lambda(u(t)-\tilde{u}(t)) \added{\in\mc{U}}$, resulting in the state transition map $\bar{x}(t,\lambda)=\phi_\mr{\bar{x}}(t,t_0,\bar{x}_0(\lambda), \bar{u}(\lambda))\in\mc{X}$. Note that this parametrization covers transitions between all possible solutions in $\mf{B}$.  Given a $\lambda$, it holds that \vskip -5mm
\begin{subequations}\label{eq:xbardyn}
\begin{align}
	\dot{\bar{x}}(t,\lambda) &= f(\bar{x}(t,\lambda),\bar{u}(t,\lambda));\\
	\bar{y}(t,\lambda) &= h(\bar{x}(t,\lambda),\bar{u}(t,\lambda)),
\end{align}
\end{subequations}
and for $\lambda\in[0,1]$, the resulting collection of trajectories
$(\bar{x}(\lambda),\bar{u}(\lambda),\bar{y}(\lambda))\in\mf{B}$ is depicted in Fig.~\ref{fig:trajectories}.
As $f,h\in \mc{C}^1$, taking the derivative of 
\eqref{eq:xbardyn} w.r.t. $\lambda$, the infinitesimal variation of the original trajectories 
{can be} analyzed. A similar concept has been introduced in \cite{CrouchSchaft1987, ForniSepulchreSchaft2013,ReyesBaesThesis2019} as \emph{variational dynamics}\footnote{In fact, we can obtain a variational system for any smooth $(x,u)$ parametrization (see \cite{ReyesBaesThesis2019} for an alternative approach).}. {Differentiation} of \eqref{eq:xbardyn} w.r.t. $\lambda$ yields the variational system \extver{\vspace{-0.1mm}}{}
\begin{equation}\label{eq:sys_diff}
	\Sigma_\delta : \left\{
	\begin{aligned}
		\delta\dot{x}(t,\lambda)&=\bar{A}(t,\lambda)\delta x(t,\lambda) + \bar{B}(t,\lambda)\delta u(t,\lambda);\\ 
		\delta y(t,\lambda) &= \bar{C}(t,\lambda)\delta x(t,\lambda) + \bar{D}(t,\lambda)\delta u(t,\lambda);
	\end{aligned}
	\right.
\end{equation}
with $\delta x(t,\lambda)= \tPartial{}{\bar{x}}{\lambda}(t,\lambda) \in\mb{R}^{n_\mr{x}}$, $\delta u(t,\lambda)= \tPartial{}{\bar{u}}{\lambda}(t,\lambda) \in\mb{R}^{n_\mr{u}}$, $\delta y(t,\lambda)= \tPartial{}{\bar{y}}{\lambda}(t,\lambda) \in\mb{R}^{n_\mr{y}}$, and, by omitting the dependence on $t$ and $\lambda$ for brevity, $\bar{A}=A(\bar{x},\bar{u}),\ldots,\bar{D}=D(\bar{x},\bar{u}) $ with 
\begin{equation}\label{eq:diffabcd}
\begin{aligned}
A(\bar{x},\bar{u}) &= \tPartial{}{f}{x}(\bar{x},\bar{u}),  &  B(\bar{x},\bar{u}) &= \tPartial{}{f}{u}(\bar{x},\bar{u}),\\
C(\bar{x},\bar{u}) &= \tPartial{}{h}{x}(\bar{x},\bar{u}),  & D(\bar{x},\bar{u}) &= \tPartial{}{h}{u}(\bar{x},\bar{u}),
\end{aligned}
\end{equation}
where $(\bar{x},\bar{u})\in\pi_{x,u}\mathfrak{B}$ for all $\lambda\in[0,\,1]$.
Note that for a trajectory in $\mf{B}$, $\lambda$ is fixed for all time. Hence, given a trajectory $\bar{x}(\lambda_*)$, corresponding to $\lambda_*$, its variations are captured in $\delta x(\lambda_*)$. In the sequel, we will generally omit $\lambda$ for brevity, e.g., $\delta x(t)= \delta x(t,\lambda_*)$.
Analogous to the primal form, solutions of the variational system \eqref{eq:sys_diff} are considered in the ordinary sense and are restricted to have left-compact support. In this paper, we will refer to \eqref{eq:sys_diff} as the \emph{differential form} of the nonlinear system \eqref{eq:sys}.

With the differential form of a system defined, we can define the notion of \emph{differential dissipativity}, interpreted as the `energy' dissipation of variations of the system trajectory that are not forced by the input. If the energy of these variations in the system trajectories decreases over time, the trajectory 
{will} eventually only be determined by the input of the system. Hence, the primal form of the system will converge to a \emph{steady-state} solution, which is not necessary a forced equilibrium point, e.g., it can be a periodic orbit. We use the definition from \cite{ForniSepulchre2013}.

\begin{definition}[Differential dissipativity]
Consider a system $\Sigma$ of the form \eqref{eq:sys} and its differential form \eqref{eq:sys_diff}, $\Sigma_\delta$. $\Sigma$ is differentially dissipative w.r.t. a supply function $\supply_\delta:\mb{R}^{n_\mr{u}}\times\mb{R}^{n_\mr{y}}\to\mb{R}$, if there exists a storage function ${\storage_\delta}: \mc{X}\times\mb{R}^{n_\mr{x}}\to\mb{R}_0^+$, with ${\storage_\delta(\cdot,0)=0}$, such that
\begin{multline}
{\storage_\delta}\big(\bar x(t_1),\delta x(t_1)\big) - {\storage_\delta}\big(\bar x(t_0),\delta x(t_0)\big) \\ \le \integral{t_0}{t_1}\supply_\delta\big(\delta u(t),\delta y(t)\big)\dif t, \label{eq:DIE_diff}
\end{multline}
for all $(\bar{x}, \bar{u})\in\pi_{x,u}\mathfrak{B}$ and for all $t_0,t_1\in\mb{R}$, with $t_0 \le t_1$.
\label{def:diffdiss}
\end{definition}
Differential \emph{passivity} definitions can be found in \cite{vanderSchaft2013, ForniSepulchreSchaft2013}. 

\begin{remark}
Note that when the incremental and differential storage functions ${\storage_\Delta}$ and ${\storage_\delta}$ are differentiable, we can also define the \emph{differentiated} forms of \eqref{eq:IDIE} and \eqref{eq:DIE_diff}.
\end{remark}

Despite the interest in general dissipativity, incremental dissipativity and differential dissipativity, the underlying connection between these notions have not been explored in the literature yet. We will establish this connection in case of quadratic supply functions in the next section, based on which performance analysis of nonlinear systems is achieved. {Furthermore, we will discuss implications of these dissipativity notions on stability 
as well.} \extver{\vspace{-1mm}}{}
\shortsection{Main results}\label{section:mainresults}
In this section, we present our main results. We first examine differential dissipativity, then we show that this property implies incremental dissipativity and general dissipativity of the nonlinear system.

\shortsubsection{Differential dissipativity of a nonlinear system}\label{section:differentialdissipativity}
Consider the differential form \eqref{eq:sys_diff} of a nonlinear system, which describes the variation of the system over a trajectory $(\bar x, \bar u, \bar y) \in \mf{B}$. Note that this system always exists if the mappings $f$ and $h$ are in $\mc{C}^1$. To formulate {our results} for differential dissipativity, we consider a quadratic storage function of the form
\begin{equation}
\storage_\delta(\bar x, \delta x) = \delta x\shrtranspose M(\bar x) \delta x,
\label{eq:storage_diff}
\end{equation}
where we assume: 
\assume{The matrix function $M\in\mc{C}^1$ is real, symmetric, bounded and positive definite, i.e., $\exists \, k_1 , k_2 \in\mb{R}^+$, such that $\forall \bar x (t)\in\mc{X}$, $k_1 I\negsemidef M(\bar x(t) ) \negsemidef k_2 I$.}{mbar}
This storage function represents the energy of the variation along the state trajectory $\bar x$. 
{We consider the following quadratic supply function,}
\begin{equation}
\supply_\delta(\delta u,\, \delta y) = \begin{pmatrix}\delta u \\ \delta y \end{pmatrix}\shrtranspose \QSR \begin{pmatrix}\delta u \\ \delta y \end{pmatrix},
\label{eq:supply_diff}
\end{equation}
with real, constant, bounded matrices $R=R^\top$, $Q=Q^\top$ and $S$. 
With \eqref{eq:storage_diff} and \eqref{eq:supply_diff}, we formulate the following theorem.
\begin{theorem}[Differential dissipativity condition]
The system in primal form \eqref{eq:sys} is differentially dissipative w.r.t. the quadratic supply function \eqref{eq:supply_diff} under a quadratic storage function \eqref{eq:storage_diff} satisfying A\ref{A:mbar}, if and only if for all $(\bar{x},\bar{u})\in\pi_{x,u}\mathfrak{B}$ and $t\in\mathbb{R}$, omitting dependence on time for brevity,
\begin{multline}
(*)\shrtranspose \! \begin{pmatrix}  \dot{M}(\bar x) &M(\bar x)\\M(\bar x)&0 \end{pmatrix} \! 
\begin{pmatrix} 
I & 0 \\ 
A\left(\bar{x},\bar{u}\right) & B\left(\bar{x},\bar{u}\right)
\end{pmatrix}  \\
-(*)\shrtranspose \QSR \begin{pmatrix} 
0 & I \\ 
C\left(\bar{x},\bar{u}\right) & D\left(\bar{x},\bar{u}\right)
\end{pmatrix} \! \preccurlyeq 0, 
\label{eqlemma1}
\end{multline}
with $\dot{M}\!(\bar x)\!=\!\!\sum_{{i=1}}^{n_\mr{x}}\!\tPartial{}{M(\bar x)}{\bar{x}_i}\dot{\bar{x}}_i$, $\dot{\bar{x}}\!=\!\differ{\bar{x}}{t}$, \mbox{and $A,\!{\scriptstyle \ldots},\!D$ as in \eqref{eq:diffabcd}.}
\extver{\vspace{-2mm}}{}
\label{theorem:differential}
\end{theorem}
\begin{proof}
By Definition~\ref{def:diffdiss}, the primal form \eqref{eq:sys} is differentially dissipative, if the differential form \eqref{eq:sys_diff} is dissipative. Hence, it suffices to show that if \eqref{eqlemma1} holds, the differential form is dissipative with storage function \eqref{eq:storage_diff} and supply function \eqref{eq:supply_diff}. Note that \eqref{eq:storage_diff} is differentiable. Therefore, we start with substituting \eqref{eq:storage_diff} and \eqref{eq:supply_diff} into the differentiated differential dissipation inequality, 
\begin{equation}\label{eq:ddpf:ddie}
\differ{}{t}\Big(\delta x\shrtranspose M(\bar x) \delta x\Big) \le \begin{pmatrix}\delta u \\ \delta y \end{pmatrix}\shrtranspose \QSR \begin{pmatrix}\delta u \\ \delta y \end{pmatrix}.
\end{equation}
By \cite{Willems1972}, \eqref{eq:ddpf:ddie} is satisfied for all possible trajectories of \eqref{eq:sys_diff} if and only if \eqref{eq:ddpf:ddie} holds for all values $(\delta x(t), \delta u(t), \delta y(t)) \in \mb{R}^{n_\mr{x}}\times \mb{R}^{n_\mr{u}}\times \mb{R}^{n_\mr{y}}$, and $\bar x(t) \in \mc{X}$. Writing out \eqref{eq:ddpf:ddie} yields,
\begin{multline}\label{eq:pfdiff:writtenout}
2\delta x\shrtranspose M(\bar x)\big(A\left(\bar{x},\bar{u}\right)\delta x + B\left(\bar{x},\bar{u}\right)\delta u\big) + \delta x\shrtranspose \dot{M}(\bar x)\delta x \le \hspace{3mm} \\ \delta u\shrtranspose Q \delta u + 2 \delta u\shrtranspose S \big(C\left(\bar{x},\bar{u}\right)\delta x + D\left(\bar{x},\bar{u}\right)\delta u\big)\\
+ \big(\!*\!\big)\shrtranspose \! R \big(C\left(\bar{x},\bar{u}\right)\delta x + D\left(\bar{x},\bar{u}\right)\delta u\big),
\end{multline}
with $A,\dots,D$ as in \eqref{eq:diffabcd} and $\dot{M}(\bar x)=\sum_{{ i=1}}^{n_\mr{x}}\tPartial{}{M(\bar x)}{\bar{x}_i}\dot{\bar{x}}_i$. It is trivial to see that \eqref{eq:pfdiff:writtenout} is equivalent to the pre- and post multiplication of \eqref{eqlemma1} with $\coldef{\delta x, \delta u}^\top$ and $\coldef{\delta x, \delta u}$, respectively. Requiring \eqref{eq:pfdiff:writtenout} to hold for all $(\bar{x},\bar{u})\in\pi_{x,u}\mathfrak{B}$ and $t\in\mathbb{R}$ is equivalent to require the condition in \eqref{eqlemma1} to hold for all $(\bar{x},\bar{u})\in\pi_{x,u}\mathfrak{B}$ and $t\in\mathbb{R}$, which proves the statement.
\end{proof}
Note that the velocity of $\bar{x}$ is required to verify differential dissipativity. Often this is solved in practice by capturing $\dot{\bar{x}}$ in a set $\mathcal{D}$, such that $\dot{\bar{x}}\in\mathcal{D}$ for all time.
%
%
%
%
\shortsubsection{Incremental dissipativity of a nonlinear system}
First, we show that the property of differential dissipativity under supply function \eqref{eq:supply_diff} implies the property of incremental dissipativity with supply function\begin{equation}
 \supply_\Delta(u,\tilde{u}, y, \tilde{y}) = \diffstate{u}{y}\shrtranspose \QSR \diffstate{u}{y}.
 \label{eq:supply_inc}
\end{equation}
Secondly, we give a computable condition to analyze incremental dissipativity. The following result is the core of our contribution.  
\begin{theorem}[Induced incremental dissipativity]\label{thm:ind-incr-dissip} When the system in primal form \eqref{eq:sys}  is differentially dissipative w.r.t.~the supply function \eqref{eq:supply_diff} with $R\negsemidef0$ under a storage function $\storage_\delta$, then there exists a storage function $\storage_\Delta$ such that the system is incrementally dissipative w.r.t.~the supply function \eqref{eq:supply_inc}.
\end{theorem}
\extver{\vspace{-1mm}}{}
\begin{proof}
By writing out the $\lambda$-dependence in \eqref{eq:DIE_diff}  for differential dissipativity, 
allows to integrate it over $\lambda$:
\begin{multline}\label{eq:jib1}
\hspace{-.5em}\integral{0}{1}\Big[\storage_\delta\big(\bar x(t_1,\lambda),\delta x(t_1,\lambda)\big)\! -\! \storage_\delta\big(\bar x(t_0,\lambda),\delta x(t_0,\lambda)\big)  -\\ \integral{t_0}{t_1}\supply_\delta\big(\delta u(\tau,\lambda),\delta y(\tau,\lambda)\big)\dif\tau\Big]\dif\lambda \le 0.
\end{multline}
We compute the integral of the storage terms first. We define the following minimum energy path between $ x$ and $\tilde x$ by
\begin{equation}\label{eq:geodesic}
\chi_{(x,\tilde x)}(\lambda):= \arginf_{\hat{x}\in\Gamma_\mathcal{X}(x,\tilde x)}\int_0^1\storage_\delta\!\left(\hat{x}(\lambda),{\Partial{}{\hat{x}(\lambda)}{\lambda}}\right)\dif\lambda.
\end{equation}
When $V_\delta(\bar x,\delta x) = \delta x^\top M(\bar x)\delta x$, $\chi_{(x,\tilde x)}$ can be seen as the geodesic connecting $x$ and $\tilde x$ corresponding to the Riemannian metric $M(\bar x)$, see also \cite{Manchester2018,ReyesBaesThesis2019}. Next, we define 
\begin{equation}\label{eq:jibstorage}
	\storage_\Delta(x,\tilde x):= \int_0^1\storage_\delta\!\left(\chi_{(x,\tilde x)}(\lambda),{\Partial{}{\chi_{(x,\tilde x)}(\lambda)}{\lambda}}\right)\dif\lambda,
\end{equation}
which will be our incremental storage function. Note that $\storage_\Delta\ge0$ as by definition $\storage_\delta\geq 0$. Furthermore, $\storage_\Delta(x,\tilde x)=0$ when $x=\tilde{x}$ as $\Partial{}{\chi_{(x,\tilde x)}(\lambda)}{\lambda}=0$ and by definition $\storage_\delta(\cdot,0)=0$. Using this incremental storage function, we have that 
\begin{equation}\label{eq:jib-1}
	\storage_\Delta(x(t_1),\tilde x(t_1))\leq \integral{0}{1}\storage_\delta\big(\bar x(t_1,\lambda),\delta x(t_1,\lambda)\big)\hspace{0.3mm}\dif\lambda,
\end{equation}
for any $\bar{x}(t_1)\in \Gamma_\mc{X}(x(t_1),\tilde x(t_1)) $ with $x(t_1),\,\tilde x(t_1)\in\mc{X}$, $t_1\in\mb{R}$, and $\bar{x}(\lambda)\in\pi_x \mf{B}$ for any $\lambda\in[0,1]$.
Furthermore, we take as parametrization for our initial condition $\bar x(t_0,\lambda) = \bar x_0(\lambda) = \chi_{(x_0,\tilde x_0)}(\lambda)$. Hence, we have that
\begin{equation}\label{eq:jib0}
\hspace{-1pt}-\storage_\Delta(x(t_0),\tilde x(t_0))\! =\! -\integral{0}{1} \storage_\delta\big(\bar x(t_0,\lambda),\delta x(t_0,\lambda)\big)\hspace{0.1mm}\dif\lambda. \hspace{-2pt}
\end{equation}
Combining \eqref{eq:jib-1} and \eqref{eq:jib0} gives that
\begin{multline*}
	\storage_\Delta\big(x(t_1),\tilde{x}(t_1)\big)-\storage_\Delta\big(x(t_0),\tilde{x}(t_0)\big)\le\\
	\integral{0}{1}\big[\storage_\delta\big(\bar x(t_1,\lambda),\delta x(t_1,\lambda)\big)\! -\! \storage_\delta\big(\bar x(t_0,\lambda),\delta x(t_0,\lambda)\big) \big]\dif\lambda.
\end{multline*}
This together with \eqref{eq:jib1} implies
\begin{multline}\label{eq:newincrem_half}
\storage_\Delta\big(x(t_1),\tilde{x}(t_1)\big)-\storage_\Delta\big(x(t_0),\tilde{x}(t_0)\big)\le\\ \integral{0}{1} \integral{t_0}{t_1}\supply_\delta\big(\delta u(\tau,\lambda),\delta y(\tau,\lambda)\big)\dif\tau\dif\lambda.
\end{multline}
We now consider the right-hand side of the inequality \eqref{eq:newincrem_half}. Changing the order of integration gives
\begin{equation}\label{eq:jib3}
\int_{t_0}^{t_1}\int_0^1(*)^\top\QSR\begin{pmatrix}\delta u(\tau,\lambda) \\ \delta y(\tau,\lambda)\end{pmatrix}\dif\lambda\,\dif\tau.
\end{equation}
We now solve the individual terms in the inner integral, 
\begin{multline}\label{eq:jib2}
\hspace{-1em}\int_0^1\!(*)\shrtranspose\!\QSR\!\begin{pmatrix}\delta u(\tau,\lambda) \\ \delta y(\tau,\lambda)\end{pmatrix}\dif\lambda = \integral{0}{1}(*)\shrtranspose Q\delta u(\tau,\lambda)\dif\lambda + \\
2\integral{0}{1}\delta u(\tau,\lambda)^\top S\delta y(\tau,\lambda)\dif\lambda +\integral{0}{1}(*)^\top R\delta y(\tau,\lambda)\dif\lambda.
 \end{multline}
Taking $\bar{u}(t,\lambda) =\tilde{u}(t)+\lambda(u(t)-\tilde{u}(t))$ as a parametrization, we obtain $\delta u(t) = \Partial{}{\bar{u}(t,\lambda)}{\lambda}=u(t)-\tilde{u}(t)$. {Hence, }the first term in \eqref{eq:jib2} resolves to $(*)^\top Q (u(\tau)-\tilde{u}(\tau))$, while the second term gives
\begin{multline}
2(u(\tau)-\tilde{u}(\tau))^\top S{\integral{0}{1} \extver{\tPartial{}{\bar{y}(\tau,\lambda)}{\lambda}}{\Partial{}{\bar{y}(\tau,\lambda)}{\lambda}}\dif\lambda}=\\
2(u(\tau)-\tilde{u}(\tau))^\top S\, ({y}(\tau) -\tilde{y}(\tau)).
\end{multline}
For the third term in \eqref{eq:jib2} where $R\negsemidef0$, i.e., $\unaryminus R\possemidef 0$, we use Lemma~\ref{lem:appendix} in Appendix \ref{sec:appendix} to obtain an upper bound:
\begin{multline}
\integral{0}{1} \extver{\big[}{\Big[} (*)^\top R\, \extver{\tPartial{}{\bar{y}(\tau,\lambda)}{\lambda}}{\Partial{}{\bar{y}(\tau,\lambda)}{\lambda}}\extver{\big]}{\Big]}\dif\lambda\le
(*)^\top\! R \left(\integral{0}{1} \extver{\tPartial{}{\bar{y}(\tau,\lambda)}{\lambda}}{\Partial{}{\bar{y}(\tau,\lambda)}{\lambda}}\dif\lambda\right) = \\
(*)^\top R\, (\bar{y}(\tau,1)-\bar{y}(\tau,0))=(*)^\top R\,(y(\tau)-\tilde{y}(\tau)).
\end{multline}
Combining our results yields
\begin{equation}\label{eq:jib4}
\int_{t_0}^{t_1}(*)^\top\QSR\begin{pmatrix}u(\tau)-\tilde{u}(\tau) \\ y(\tau)-\tilde{y}(\tau)\end{pmatrix}\dif\tau,
\end{equation}
as an upper bound for \eqref{eq:jib3}. Thus, if \eqref{eq:DIE_diff} holds, we know that \eqref{eq:jib1} holds, which in turn implies, considering a supply function \eqref{eq:supply_diff} with $R\negsemidef 0$, that 
\begin{multline}\label{eq:jib5}
\storage_\Delta\big(x(t_1),\tilde{x}(t_1)\big)-\storage_\Delta\big(x(t_0),\tilde{x}(t_0)\big)\le\\ \int_{t_0}^{t_1}(*)^\top\QSR\begin{pmatrix}u(\tau)-\tilde{u}(\tau) \\ y(\tau)-\tilde{y}(\tau)\end{pmatrix}\dif\tau,
\end{multline}
via the upper bound \eqref{eq:jib4}.
Hence, if the system is differentially dissipative w.r.t. the supply function \eqref{eq:supply_diff} with $R\negsemidef0$, then the system is incrementally dissipative w.r.t. the equally parametrized supply function \eqref{eq:supply_inc}.
\end{proof}
\begin{remark}[Restricted $R$]
Restriction $R\negsemidef0$ is a technical necessity in the proof of Theorem \ref{thm:ind-incr-dissip}. In case of $R\posdef0$ or $R$  being indefinite, validity of Theorem \ref{thm:ind-incr-dissip} is an open question.
\end{remark}
Comparing Theorem \ref{thm:ind-incr-dissip} to existing results in this context, we want to highlight that \cite{WaitmanPaper2018, waitman2016incremental} also give some results on incremental dissipativity. However, these works only focus on a specific and restrictive form of the supply function. Moreover, the technical result of \cite{WaitmanPaper2018} refers to a proof in a paper that has never appeared to the authors' knowledge.

From Theorem \ref{thm:ind-incr-dissip}, we have the following (trivial) result:
\begin{corollary}[Incremental dissipativity condition]\label{coro:incrementaldissip} The system in primal form \eqref{eq:sys} is incrementally dissipative w.r.t. the supply function \eqref{eq:supply_inc} with $R\negsemidef0$, if \eqref{eqlemma1} holds for all $(\bar{x},\bar{u})\in\pi_{x,u}\mathcal{B}$ with $M$ satisfying A\ref{A:mbar}.
\end{corollary}
Corollary \ref{coro:incrementaldissip} gives a sufficient condition to verify incremental dissipativity of a general nonlinear system. Note that by this result, if the matrix inequality \eqref{eqlemma1} holds for all $(\bar x, \bar u)\in\pi_{x,u}\mathcal{B}$, then we know that there exist a valid storage function of the form \eqref{eq:jibstorage}. However, calculating this function in an explicit form might be difficult (see Section \ref{sec:inc:sto}). If no positive definite $M$ can be found to satisfy \eqref{eqlemma1}, then it does \emph{not} necessarily mean that the system is not differentially or incrementally dissipative. Inequality \eqref{eqlemma1} might hold for a non-quadratic ${\storage_\delta}$, or a more complex $M$.
\shortsubsection{Explicit incremental storage function}\label{sec:inc:sto}
{Even if deriving an} explicit form of 
 \eqref{eq:jibstorage} {is challenging in general}, under the quadratic form of \eqref{eq:storage_diff}, we can take {an} extra assumption to give an explicit construction:
\assume{$M(\bar x)$ can be decomposed as $M(\bar x)=N^{\top}\!(\bar{x}) PN(\bar{x})$, $P\posdef 0$, and $\exists\, \nu:\mb{R}^{n_\mr{x}}\!\to\!\mb{R}^{n_\mr{x}}$ s.t. $\tPartial{}{\nu(\bar{x})}{x}=N(\bar{x})$.}{mdiff}
\added{While this decomposition of $M(\bar{x})$ is always possible if it satisfies A\ref{A:mbar}, see \cite{decompdoc}, existence of $\nu$ such that $\tPartial{}{\nu(\bar{x})}{x}=N(\bar{x})$ is not guaranteed for any $M(\bar{x})$}. This illustrates well the {challenges for obtaining} an explicit {construction of} $\storage_\Delta$. 
For the sake of simplicity, we assume in the remainder of this subsection that $\mathcal{X}=\mathbb{R}^{n_\mr{x}}$. 
\begin{lemma}[Induced \hspace{-.15ex}incremental \hspace{-.15ex}storage \hspace{-.15ex}function] \label{lem:nine} 
\hspace{0em} If the system in primal form \eqref{eq:sys} is differentially dissipative with a 
storage function 
\begin{equation*}
	\storage_\delta(\bar x, \delta x) = \delta x\shrtranspose M(\bar x) \delta x, 
\end{equation*}
where $M$ satisfies A\ref{A:mbar} {and  A\ref{A:mdiff}}, then the incremental storage function $\storage_\Delta$ in Theorem \ref{thm:ind-incr-dissip} is given by
	 \begin{equation}\label{eq:incrstoragefun}
	 	\storage_\Delta(x,\tilde{x}) = (\nu(x)-\nu(\tilde x))\shrtranspose P(\nu(x)-\nu(\tilde x)).
	 \end{equation}
	Additionally, if $M(\bar x)=M$ for all $\bar{x}\in\mc{X}$, then, the incremental storage function simplifies to
	  \begin{equation}\label{eq:storage_incremental_constant}
	 	\storage_\Delta(x,\tilde{x}) = (x-\tilde x)\shrtranspose M(x-\tilde x).
	 \end{equation}
\end{lemma}
\extver{\vspace{-2mm}}{}
\begin{proof}
	Based on \eqref{eq:jib1}, we need to compute the terms
	\begin{subequations}
	\begin{equation}\label{eq:intstorage1}
		\integral{0}{1} \storage_\delta(\bar{x}(t_1,\lambda),\delta x(t_1,\lambda))\dif \lambda 	
		\end{equation}
	and
	\begin{equation}\label{eq:intstorage2}
		-\integral{0}{1} \storage_\delta(\bar{x}(t_0,\lambda),\delta x(t_0,\lambda))\dif \lambda.
	\end{equation}
	\end{subequations}
	Based on A\ref{A:mdiff}, we can decompose $M(\bar{x})$ into
\begin{equation}\label{eq:Mdecompose}
	M(\bar{x}) = N^\top\!(\bar{x})P N(\bar{x}),
\end{equation}
where $P\posdef 0$ with $P\in\mathbb{R}^{n_\mr{x}\times n_\mr{x}}$ and, because of A\ref{A:mbar}, $N(\bar{x}(t,\lambda))\in\mathbb{R}^{n_\mr{x}\times n_\mr{x}} $ is invertible on $\mc{X}$, i.e., $\det{N(\bar{x})}\neq 0, \,\forall\,\bar{x}\in\mc{X}$. {Furthermore, by A\ref{A:mdiff}, there exists a  diffeomorphism $\nu:\mathbb{R}^{n_\mr{x}}\rightarrow\mathbb{R}^{n_\mr{x}}$ such that $\Differ{}{\nu}{x}(\bar{x})=N(\bar{x}),\,\forall\,\bar{x}\in\mc{X}$.} Next, define 
	$\bar{z}(t,\lambda) := \nu(\bar{x}(t,\lambda))$,
which satisfies that
\begin{equation}
	\delta z(t,\lambda) = \tPartial{}{}{\lambda}\bar{z}(t,\lambda) = N(\bar{x}(t,\lambda))\delta x(t,\lambda).
\end{equation}
This allows to rewrite \eqref{eq:storage_diff} as 
\begin{equation}\label{eq:deltaxtodeltaz}
\delta x^\top\! M(\bar{x}) \delta x = \delta x^\top \!N^\top\!(\bar{x}) P  N(\bar{x}) \delta x = \delta z^\top\! P \delta z.
\end{equation}
Using this relation,  
the first term \eqref{eq:intstorage1} can be written as
\begin{equation}\label{eq:intstorage1_z}
	\integral{0}{1} \delta z\shrtranspose\hspace{-0.2mm}(t_1,\lambda) P \delta z(t_1,\lambda)\dif \lambda.
\end{equation}
	Applying Lemma \ref{lem:appendix}, see Appendix \ref{sec:appendix}, to \eqref{eq:intstorage1_z} results in 
	\begin{equation*}
		(*)^\top P \left(\integral{0}{1} \delta z(t_1,\lambda)\dif\lambda\right)\le \integral{0}{1} (\delta z\shrtranspose\hspace{-0.2mm}(t_1,\lambda) P \delta z(t_1,\lambda))\dif \lambda.
	\end{equation*}
	Hence,
\begin{multline}\label{eq:intstorage1rewritten}
\hspace{-5mm}
\left(*\right)^{\!\top}\!\! P\bigl(\nu(x(t_1))-\nu(\tilde{x}(t_1))\bigr)=
\left(*\right)^{\!\top}\!\! P\bigl(\bar{z}(t_1,1)-\bar{z}(t_1,0)\bigr)\le\\
\integral{0}{1}(\delta z\shrtranspose\hspace{-0.2mm}(t_1,\lambda) P\delta z(t_1,\lambda))\dif\lambda = \\
 \hspace{2mm}\integral{0}{1}\big(\delta x\shrtranspose\hspace{-0.2mm}(t_1,\lambda) M(\bar{x}(t_1,\lambda))\delta x(t_1,\lambda)\big)\dif\lambda.\hspace{-2mm}
\end{multline}
Before looking at the second term, i.e., \eqref{eq:intstorage2}, let us recall some definitions. As aforementioned, the parametrized initial condition $\bar x(t_0,\lambda)=\bar x_0(\lambda)$ can be taken as any smooth parametrization {$\bar x_0\in\Gamma_\mathcal{X}(x_0,\tilde{x}_0)$}. 
{Recall that $\nu$ is a diffeomorphism, implying that $\nu^{-1}:\mathbb{R}^{n_\mr{x}}\rightarrow\mathbb{R}^{n_\mr{x}}$ exists and  $\nu,\nu^{-1}\in\mc{C}_1$.}
Hence, w.l.o.g. we take
\begin{equation}
	\bar{x}_0(\lambda) = \nu^{-1}(\bar{z}(t_0,\lambda)), \label{eq:para}
\end{equation}
where
$
	\bar{z}(t_0,\lambda) = \nu(\tilde x_0)+\lambda(\nu(x_0)-\nu(\tilde{x}_0)). 
$ 
Note that this choice of $\bar x_0(\lambda)$ satisfies the aforementioned conditions. 
Consequently, we have that 
\begin{equation}\label{eq:deltazic}
	\delta z(t_0,\lambda) = \tPartial{}{}{\lambda}\bar{z}(t_0,\lambda) = \nu(x_0)-\nu(\tilde x_0).
\end{equation}
Using this result and \eqref{eq:deltaxtodeltaz}, the second term \eqref{eq:intstorage2} gives
\begin{multline}\label{eq:intstorage2rewritten}
		\integral{0}{1}\left(\nu(x_0)-\nu(\tilde x_0)\right)\shrtranspose P \left(\nu(x_0)-\nu(\tilde x_0)\right)\dif \lambda =\\ \left(\nu(x_0)-\nu(\tilde x_0)\right)\shrtranspose P \left(\nu(x_0)-\nu(\tilde x_0)\right).
\end{multline}
Combining the results of \eqref{eq:intstorage1rewritten} and \eqref{eq:intstorage2rewritten}, it holds that
\begin{multline}
	\hspace{-1em}\integral{0}{1} \big[\storage_\delta(\bar{x}(t_1,\lambda),\delta x(t_1,\lambda))- \storage_\delta(\bar{x}(t_0,\lambda),\delta x(t_0,\lambda))\big]\dif \lambda\ge \\
	(\ast)^\top P\left(\nu(x(t_1))-\nu(\tilde{x}(t_1))\right)-\\(\ast)\shrtranspose P \left(\nu( x(t_0))-\nu(\tilde  x(t_0))\right) .
\end{multline}
Combining this result with \eqref{eq:jib4} gives
\begin{multline}
	\storage_\Delta(x(t_1),\tilde x(t_1))-\storage_\Delta(x(t_0),\tilde x(t_0))\leq\\ \int_{t_0}^{t_1}(*)^\top\QSR\begin{pmatrix}u(\tau)-\tilde{u}(\tau) \\ y(\tau)-\tilde{y}(\tau)\end{pmatrix}\dif\tau,
\end{multline}
where $\storage_\Delta$ is according to \eqref{eq:incrstoragefun}. Hence, \eqref{eq:incrstoragefun} qualifies as an incremental storage function for \eqref{eq:sys}.

In case  $M(\bar x)=M$ for all $\bar{x}\in\mc{X}$, the decomposition in \eqref{eq:Mdecompose} simplifies to $N=I$ and $P=M$, hence, $\nu(x) = x$ and we obtain \eqref{eq:storage_incremental_constant}. Note that the same result is obtained when solving \eqref{eq:geodesic} and \eqref{eq:jibstorage} directly for $V_\delta(\bar x,\delta x) = \delta x^\top M \delta x$, as in that case $\chi_{(x,\tilde x)} = \tilde x+\lambda (x-\tilde x)$ and hence $V_\Delta$ is given by \eqref{eq:storage_incremental_constant}.
\end{proof}
\vskip -2mm

\begin{figure}
\centering
\includegraphics[scale=0.7,trim={0 3.5mm 0 0},clip]{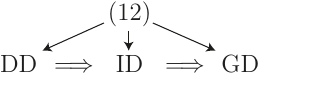}
\extver{\vspace{-2mm}}{}
\caption{Chain of implications with the dissipativity notions: differential dissipativity (DD) implies incremental dissipativity (ID), and incremental dissipativity implies general dissipativity (GD). Condition \eqref{eqlemma1} can be used to analyze the various dissipativity notions.}\label{fig:chain}
\extver{\vspace{-2mm}}{}
\end{figure}

{In case $\mathcal{X}$ is a bounded convex set, Lemma \ref{lem:nine} can be also shown to hold true, if either beyond  A\ref{A:mdiff} it holds that $\nu(\mathcal{X})$ is also convex, or if $M$ is a constant matrix. }
\shortsubsection{General dissipativity analysis of a nonlinear system}\label{section:linkincrementaldifferential}
We now show that incremental dissipativity implies that the considered system is globally dissipative, i.e., dissipative w.r.t. any forced equilibrium point in $\mathcal{B}$. 
\begin{theorem}[Induced general dissipativity]\label{theorem:incr_normal}
Given a nonlinear system in its primal form \eqref{eq:sys}. Suppose that $(x_\mr{e},u_\mr{e}, y_\mr{e})\in \mathcal{B}$ is a (forced) equilibrium point of the system, i.e., $(\breve{x}(t), \breve{u}(t), \breve{y}(t)) = (x_\mr{e},u_\mr{e}, y_\mr{e})$ and $(\breve{x}(t), \breve{u}(t), \breve{y}(t))$ satisfies \eqref{eq:sys} for all $t\in\mathbb{R}$.
If the system is incrementally dissipative under the supply function \eqref{eq:supply_inc}, then for every equilibrium $(x_\mr{e},u_\mr{e}, y_\mr{e})$, the system is dissipative w.r.t. an equally parametrized  supply function.
\end{theorem}
\begin{proof}
If the system is incrementally dissipative w.r.t. the supply function \eqref{eq:supply_inc} under the storage function $\storage_\Delta$, then it holds that
\begin{multline*}
\storage_\Delta\big(x(t_1), \tilde{x}(t_1)\big) - \storage_\Delta\big(x(t_0), \tilde{x}(t_0)\big) \le\\ \int_{t_0}^{t_1}(*)\shrtranspose  \QSR \begin{pmatrix}
	u(t)-\tilde u(t)\\y(t)-\tilde y(t)
\end{pmatrix}\dif t,
\end{multline*}
for all $t_0,t_1\in\mb{R}$, with $t_0 \le t_1$.
Let the trajectory $(\tilde{x},\tilde{u}, \tilde{y})$ be equal to the equilibrium trajectory $(\breve{x},\breve{u}, \breve{y})$, i.e., the equilibrium point $(x_\mr{e},u_\mr{e}, y_\mr{e})$. Hence, for all $t_0,t_1\in\mb{R}$, with $t_0 \le t_1$
\begin{multline*}
\storage_\Delta\big(x(t_1), x_\mr{e}\big) - \storage_\Delta\big(x(t_0), x_\mr{e}\big) \le\\ \int_{t_0}^{t_1}(*)\shrtranspose  \QSR \begin{pmatrix}
	u(t)-u_\mr{e}\\y(t)-y_\mr{e}
\end{pmatrix}\dif t,
\end{multline*}Next,  introduce the coordinate shift
\begin{equation*}
q =x-x_\mr{e},\quad\! w=u-u_\mr{e}, \quad\! z = y-y_\mr{e},
\end{equation*}
and define
\begin{equation}\label{eq:gendissipfromincrstorage}
	\storage(q) := \storage_\Delta(q+x_\mr{e},x_\mr{e}),
\end{equation}
which is non-negative and satisfies that $\storage(0)=0$.
Substituting this in the inequality gives that
\begin{equation*}
\storage\big(q(t_1)\big) - \storage\big(q(t_0)\big) \le \int_{t_0}^{t_1}(*)\shrtranspose  \QSR \begin{pmatrix}
	w(t)\\z(t)
\end{pmatrix},
\end{equation*} holds for all $t_0,t_1\in\mb{R}$, with $t_0 \le t_1$, which is the general dissipation inequality \eqref{eq:DIE} with $\storage$ as defined in \eqref{eq:gendissipfromincrstorage} being the corresponding storage function. Hence,  \eqref{eq:sys}  is dissipative w.r.t. any arbitrary forced equilibrium point if it is incrementally dissipative.
\end{proof}
\extver{\vspace{-2mm}}{}
By this last result, we have obtained a \emph{chain of implications}, which connect the notions of dissipativity. Moreover, we gave a condition (matrix inequality \eqref{eqlemma1}) that allows to examine differential, incremental and general dissipativity and thus examine global stability and performance of a nonlinear system. This chain of implications is summarized in Fig. \ref{fig:chain}.
{A result similar to Theorem \ref{theorem:incr_normal} is given in \cite{liu2014incremental} for single-input-single-output networked nonlinear systems. However, note that Theorem \ref{theorem:incr_normal} is more general, as it holds for general nonlinear multi-input-multi-output systems of the form \eqref{eq:sys}}.
%
\begin{remark}\label{rem:stab}
If the supply function satisfies
\begin{equation}\label{eq:supply_stab}
\supply(0,y)\le 0 \quad \forall  y\in\pi_y\mf{B},
\end{equation}
then {it is well-known that} dissipativity implies {Lyapunov} stability of {\eqref{eq:sys}} \cite{L2PassiveSchaft2017, angeli2002lyapunov}. {Under a similar condition on $\supply_\Delta$,} incremental dissipativity implies incremental stability, {which means that} 
{there exists a function $\beta\in\mc{KL}$, such that, for all $u\in\pi_u\mf{B}$, all $x_0,\tilde x_0\in\mc{X}$ and all $t\ge 0$,
\[
	\norm{\phi_\mr{x}(t,0,x_0,u)-\phi_\mr{x}(t,0,\tilde x_0,u)}\leq \beta(\norm{x_0-\tilde x_0},t).
\]
See \cite{angeli2002lyapunov} for more details. 
{Similarly, we have that} differential dissipativity implies stability of \eqref{eq:sys_diff} when $\forall \delta y\in{{\left(\mb{R}^{n_\mr{y}}\right)}^\mb{R}}:\supply_\delta(0,\delta y)\le 0$.} As 
$R\negsemidef 0$, {these conditions are trivially satisfied by our considered supply functions and through Theorem \ref{thm:ind-incr-dissip} and \ref{theorem:incr_normal}, the same chain of implications hold between these stability notions as in Figure \ref{fig:chain}. Hence, by showing differential dissipativity with the considered supply functions, we also show incremental and Lyapunov stability of \eqref{eq:sys}}. If the above conditions on $\supply$ hold in the strict sense, then the implications hold in terms of asymptotic forms of stability.
\end{remark}
%
%
%
%
%
%
%
\shortsection{Performance analysis via convex tests}\label{section:applications} 
We now use the dissipativity results of Section \ref{section:mainresults} to recover incremental notions of well-known performance indicators (\ltwo-gain, \linf-gain, passivity and the generalized \htwo-norm) and propose a method that allows for global, convex performance analysis of nonlinear systems. This contribution can also serve as a stepping stone for the formulation of incremental controller synthesis {methods}. We want to highlight that the results in this section resemble to conditions of respective performance indicators\extver{\footnote{Therefore, the proofs in this section are omitted. For the interested reader, the proofs are included in the extended version of this paper \cite{VerhoekKoelewijnToth2020}.}}{} of LPV systems. The LPV conditions differ from these results as we use the differential form of a nonlinear system. Hence, the relations follow from a completely different analysis {that allows for global}  
nonlinear \emph{performance analysis}. 

We will introduce the incremental performance notions for storage functions of the form {of} \eqref{eq:storage_incremental_constant}. It is trivial to extend these results to the case when a matrix function $M(\bar{x})$ is considered. 
\shortsubsection{Incremental \ltwo-gain}
A system has finite \ltwo-gain $\gamma < \infty$ if the system is dissipative w.r.t. to the supply rate $\supply(u,y) = \gamma^2 \norm{u}_2^2 - \norm{y}_2^2$ \cite{LMICourse}, i.e., $u$ must be in $\ms{L}_2^{n_\mr{u}}$. Let $\mf{B}_2$ be defined as $\mf{B}_2:=\setdefinition{(x,u,y)\in\mf{B}}{u\in\ms{L}_2^{n_\mr{u}}}$. 
There are several definitions in the literature that extend the classical \ltwo-gain definition towards the incremental setting \cite{L2PassiveSchaft2017, FromionScorlettiFerreres, KoelewijnTothMazzoccanteNijmeijer2019}. The following definition fits with the incremental dissipativity notion discussed in this paper.

\begin{definition}[Incremental \ltwo-gain]\hspace{-0pt}
The incremental \ltwo-gain, i.e., $\litwo$-gain, of the system $\Sigma$ of the form \eqref{eq:sys}  is \extver{\vspace{-0.1mm}}{}
\begin{equation}
\norm{\Sigma}_{\litwo} \!\! := \!\!\sup_{0<\norm{u-\tilde{u}}_2<\infty}\!\!\frac{\norm{y-\tilde{y}}_2}{\norm{u-\tilde{u}}_2}, \label{eq:L2normdef}
\end{equation}
where $(x,u,y),(\tilde{x},\tilde{u},\tilde{y})\in\mathfrak{B}_2$ are any two arbitrary trajectories of $\Sigma$ {for which $x(0)=\tilde{x}(0)$.} 
\label{DefinitionL2}
\end{definition}
\begin{remark}[$\litwo$-gain in the LTI case]
The \ltwo-gain and the \litwo-gain are equivalent for LTI systems \cite{PatricksHinf}. Hence, the \ltwo-gain of a differential LTI system is equal to the \ltwo-gain of a primal LTI system. 
\end{remark}
The results in \cite{L2PassiveSchaft2017, FromionScorlettiFerreres, KoelewijnTothMazzoccanteNijmeijer2019}, together with Corollary \ref{coro:incrementaldissip} lead to the following result:
\begin{corollary}[$\litwo$-gain bound]
Consider $\Sigma$ as the system \eqref{eq:sys} 
and let $\gamma\in \mathbb{R}^+$. If there exists an $ M=M^\top \posdef 0$ s.t. $\forall (\bar{x},\bar{u})\in\pi_{x,u}\mathcal{B}$, 
\begin{align}
\begin{pmatrix}
A(\bar{\eta})^\top M + M A(\bar{\eta}) & MB(\bar{\eta}) & C(\bar{\eta})^\top \\ B(\bar{\eta})^\top M &-\gamma^2I &D(\bar{\eta})^\top \\ C(\bar{\eta}) & D(\bar{\eta}) & -I
\end{pmatrix} &\preccurlyeq 0, \label{eq:L2convex}
\end{align}
where $\bar{\eta}= \coldef{\bar{x}, \bar{u}}$, then $\|\Sigma\|_{\litwo}\!\!\le\gamma$.
\label{lemma:l2}
\end{corollary}
\begin{proof}
\extver{
The proof follows by combining Corollary \ref{coro:incrementaldissip} with the results in\cite{L2PassiveSchaft2017}. See also \cite{VerhoekKoelewijnToth2020}.
}{
The proof can be found in Appendix \ref{appendix:l2}. 
}
\end{proof}
In \cite{FromionScorletti2003}, it is shown that $\|\Sigma_\delta\|_{\ltwo}<\gamma\Leftrightarrow\|\Sigma\|_{\litwo}<\gamma$. It is an interesting (open) question how necessity can also be established via Theorem \ref{thm:ind-incr-dissip} in this case. Additionally, note that \eqref{eq:L2convex} is linear, i.e., convex, in $M$ and $\gamma^2$, but it is an infinite semi-definite problem.  
We will discuss in Section \ref{subsection:towardLPV} how to turn it into a finite number of \emph{linear matrix inequalities} (LMIs)-based optimization problem.  
\shortsubsection{Incremental \linf-gain} \noindent
The well-known \lone-norm is defined for stable LTI systems that map inputs with bounded amplitude to outputs with bounded amplitude. 
For LTI systems, the \lone-norm is equivalent with the induced \linf-norm, i.e., the peak-to-peak gain of a system. We extend the notion of the \linf-gain {to the incremental setting, which characterizes} the peak-to-peak gain between two arbitrary trajectories of a system. Let $\mf{B}_\infty$ be defined as $\mf{B}_\infty:=\setdefinition{(x,u,y)\in\mf{B}}{u\in\ms{L}_\infty^{n_\mr{u}}}$. 
\begin{definition}[Incremental \linf-gain]\hspace{-0pt}
The incremental \linf-gain, i.e., \liinf-gain, of the system $\Sigma$ of the form \eqref{eq:sys} is 
\begin{equation}
\norm{\Sigma}_{\liinf} \!:= \!\!\sup_{0<\norm{u-\tilde{u}}_\infty<\infty}\!\!\frac{\norm{y-\tilde{y}}_\infty}{\norm{u-\tilde{u}}_\infty},
\end{equation}
where $(x,u,y),(\tilde{x},\tilde{u}, \tilde{y})\in\mf{B}_\infty$ are any two arbitrary trajectories of $\Sigma$ {for which $x(0)=\tilde{x}(0)$.} 
\label{def:incrLinf}
\end{definition}
As an extension of \cite[Sec.~10.3]{SchererL1norm} and \cite[Sec.~3.3.5]{LMICourse}, the following result gives a sufficient condition for an upper bound $\gamma$ of the \liinf-gain of a nonlinear system.

\begin{corollary}[\liinf-gain bound]
Consider $\Sigma$ as the system \eqref{eq:sys}  
and let $\gamma\in \mathbb{R}^+$. 
If there exist an $M\shrtranspose = M\succ 0$, $\kappa > 0$ and $\mu>0$ such that $\forall (\bar{x},\bar{u})\in\pi_{x,u}\mathcal{B}$, 
\begin{subequations}
\begin{gather}
\hspace{-3mm}\begin{pmatrix}
A\left(\bar{\eta}\right)\shrtranspose M + M A\left(\bar{\eta}\right) + \kappa M & M B\left(\bar{\eta}\right) \\
B\left(\bar{\eta}\right)\shrtranspose M&- \mu I  \end{pmatrix}  \prec 0,\hspace{-1mm} \label{eq:L1LMI1} \\
\begin{pmatrix} \kappa M & 0 & C \left(\bar{\eta}\right)\shrtranspose \\ 0 & (\gamma-\mu)I & D\left(\bar{\eta}\right)\shrtranspose \\ C\left(\bar{\eta}\right) & D\left(\bar{\eta}\right)& \gamma I \end{pmatrix}  \succ 0, \label{eq:L1LMI2}
\end{gather}
\end{subequations}
where $\bar{\eta}= \coldef{\bar{x}, \bar{u}}$, then $\|\Sigma\|_{\liinf}\!<\gamma$.
\label{corollary:l1}
\end{corollary}
\begin{proof}
\extver{
The proof follows by combining Corollary \ref{coro:incrementaldissip} with the results in\cite{SchererL1norm,LMICourse}. See also \cite{VerhoekKoelewijnToth2020}.
}{
The proof can be found in Appendix \ref{appendix:l1}.
}
\end{proof}
Despite of the fact that \eqref{eq:L1LMI1} is not convex in $\kappa$ and $M$ due to their multiplicative relation, by fixing $\kappa$ and performing a line-search over it,   \eqref{eq:L1LMI1} again corresponds to an infinite \emph{Semi-Definite Program} (SDP). 

\shortsubsection{Incremental passivity}\label{section:pass}
Passivity is a widely studied system property and it has been recently extended towards the incremental setting \cite{PavlovMarconi2007,L2PassiveSchaft2017} and the differential setting \cite{vanderSchaft2013, ForniSepulchre2013, ForniSepulchreSchaft2013}. In \cite{BeneluxPaper}, the connection between differential and incremental passivity has been established {for a storage function \eqref{eq:storage_diff} with constant $M$}. That work might serve as a parallel proof for Theorem \ref{thm:ind-incr-dissip}, when focusing only on passivity.

A system is said to be passive if it is dissipative w.r.t. to the supply rate $\supply(u,y) =  u\shrtranspose y + y\shrtranspose u$. Based on \cite{L2PassiveSchaft2017},  the definition of incremental passivity is as follows:
\begin{definition}[Incremental passivity]
A system of the form \eqref{eq:sys} is incrementally passive, if for the supply 
\begin{equation}
\supply_\Delta(u,\tilde{u},y, \tilde{y}) =  (u-\tilde{u})\shrtranspose(y- \tilde{y}) + (y- \tilde{y})\shrtranspose(u-\tilde{u}), \label{eq:defpassive}
\end{equation}
there exist a storage function $\storage_\Delta:\mathcal{X}\times\mathcal{X}\to \mathbb{R}^+$ s.t.
\begin{equation*}
\storage_{\!\Delta}\big(x(t_1),\tilde x (t_1)\big) - \storage_{\!\Delta}\big(x(t_0),\tilde x (t_0)\big) \le {\textstyle\int_{t_0}^{t_1}\,} \supply_\Delta(u,\tilde{u},y, \tilde{y})\dif t
\end{equation*}
for all $(x,u,y),(\tilde{x},\tilde{u}, \tilde{y})\in\mathfrak{B}$.
\end{definition}
Based on Corollary \ref{coro:incrementaldissip}, the following result holds:
\begin{corollary}[Incremental passivity condition]\label{cor:pass}
The system \eqref{eq:sys} with $n_\mr{y}=n_\mr{u}$ is incrementally passive if there exists an $M^\top=M\posdef0$ such that $\forall (\bar{x},\bar{u})\in\pi_{x,u}\mathcal{B}$
\begin{equation}\label{eq:condpass}
	\begin{pmatrix}
		A(\bar{\eta})^\top M + MA(\bar{\eta}) & M B(\bar{\eta}) - C(\bar{\eta})^\top \\ 
		B(\bar{\eta})^\top M - C(\bar{\eta}) & -D(\bar{\eta})-D(\bar{\eta})^\top
	\end{pmatrix}\negsemidef 0,
\end{equation}
where $\bar{\eta}=\coldef{\bar x, \bar u)}$.
\label{corollary:passive}
\end{corollary}
\begin{proof}
The proof follows by direct application of Corollary \ref{coro:incrementaldissip} with $Q=R=0$ and $S=I$.
\end{proof}
Comparing Corollary \ref{cor:pass} to \cite{BeneluxPaper} and \cite{vanderSchaft2013}, these papers give results on differential passivity for a combined primal and differential system formulation (a prolonged system \cite{CrouchSchaft1987}) using a specific form of storage function. The result depends on equality constrains, which serve as a decoupling condition between the differential storage and the primal storage, while in this paper the differential storage and the primal storage have the same structure (quadratic form with the same $M$), not requiring such equality constraints.

\shortsubsection{Generalized incremental \htwo-norm} \noindent
There are several extensions of the \htwo-norm for nonlinear systems embedded as LPV systems \cite{SouzaH2,XieH2,BoualiH2}. In this paper, we extend the notion of the generalized \htwo-norm to the incremental setting:
\begin{definition}[Incremental \htwo-norm]\label{def:h2}
Consider $\Sigma$ as the system \eqref{eq:sys} with $\Partial{}{h}{u}=0$. The generalized incremental \htwo-norm, i.e., \hitwo-norm, of $\Sigma$ is 
\begin{equation}
\big\|\Sigma\big\|_{\hitwo} \! := \!\!\sup_{0<\norm{u-\tilde{u}}_2<\infty}\!\!\frac{\norm{y-\tilde{y}}_\infty}{\norm{u-\tilde{u}}_2\hfill},
\end{equation}
where $(x,u,y),(\tilde{x},\tilde{u},\tilde{y})\in\mathfrak{B}_2$ are any two arbitrary trajectories of $\Sigma$ {for which $x(0)=\tilde{x}(0)$.} 
\end{definition}
Note that if assumption $\Partial{}{h}{u}=0$ does not hold, then the $\hitwo$-norm is trivially unbounded. As an extension of \cite[Sec.~3.3.4]{LMICourse}, the following result characterizes an upper bound $\gamma$ on the \hitwo-norm.
\begin{corollary}[$\hitwo$-gain bound]
Consider $\Sigma$ as the system \eqref{eq:sys} with $\Partial{}{h}{u}=0$ and let $\gamma\in \mathbb{R}^+$. If there exists an $M\shrtranspose = M\posdef 0$ such that $\forall (\bar{x},\bar{u})\in\pi_{x,u}\mathcal{B}$,
\extver{%
\begin{equation*}
\begin{pmatrix}
A(\bar{\eta})\shrtranspose M + M A(\bar{\eta}) & M B(\bar{\eta}) \\
B(\bar{\eta})\shrtranspose M&- \gamma I  \end{pmatrix} \! \prec 0, \>
\begin{pmatrix}
M & C (\bar{\eta})\shrtranspose \\ C(\bar{\eta}) & \gamma I\end{pmatrix} \! \succ 0,
\end{equation*}
}{%
\begin{subequations}
\begin{gather}
\begin{pmatrix}
A(\bar{\eta})\shrtranspose M + M A(\bar{\eta}) & M B(\bar{\eta}) \\
B(\bar{\eta})\shrtranspose M&- \gamma I  \end{pmatrix}  \prec 0, \label{eq:H2LMI1} \\
\begin{pmatrix}
M & C (\bar{\eta})\shrtranspose \\ C(\bar{\eta}) & \gamma I\end{pmatrix}  \succ 0, \label{eq:H2LMI2}
\end{gather}
\end{subequations}
}%
where $\bar{\eta}=\coldef{\bar x, \bar u}$, then $\|\Sigma\|_{\hitwo}\!<\gamma$. 
\label{corollary:h2}
\end{corollary}
\begin{proof}
\extver{
The proof follows by combining Corollary \ref{coro:incrementaldissip} with the results in\cite{LMICourse}. See also \cite{VerhoekKoelewijnToth2020}.
}{
The proof can be found in Appendix \ref{appendix:h2}.
}
\end{proof}
\extver{}{
\begin{remark}
Note that the performance measures discussed in this section have a specific $(Q,S,R)$-triplet associated with them. Specifically, $(Q,S,R)=(\gamma^2 I, 0, \unaryminus I)$ for \litwo-gain, $(Q,S,R)=(\mu I, 0, 0)$ for \liinf-gain,  $(Q,S,R)=(0, I, 0)$ for incremental passivity and $(Q,S,R)=(\gamma I, 0, 0)$ for the \hitwo-norm. 
\end{remark}
}
\shortsubsection{Convex computation with {D}PV inclusions}\label{subsection:towardLPV}
So far, the obtained results have yielded matrix inequalities that correspond to infinite dimensional SDPs. 
This section presents a convexification of the constraint variation to recast these problems as regular SDPs by embedding of the differential form of the system in a {D}PV inclusion. Inspired by \cite{Toth2010,WangTothManchester2020}, we define the {D}PV inclusion of \eqref{eq:sys} as follows. 
\begin{definition}[{D}PV inclusion]
The {D}PV inclusion of \eqref{eq:sys}, given by
\begin{equation}\label{eq:sys_LPV}
	\Sigma_{\mr{{D}PV}} : \left\{
	\begin{aligned}
		\delta\dot{x}(t)&=A(\sche(t))\delta x(t) + B(\sche(t))\delta u(t),\\ 
		\delta y(t) &= C(\sche(t))\delta x(t) + D(\sche(t))\delta u(t),
	\end{aligned}
	\right.
\end{equation}
with $\sche(t)\in\mathcal{P}$ being the scheduling variable, is an embedding of the differential form of \eqref{eq:sys} on the compact convex region $\mc{P}\subset\mathbb{R}^{n_\mathrm{\sche}}$, if there exists a function $\psi : \mathbb{R}^{n_{{\mathrm{x}}}} \times \mathbb{R}^{n_{{\mathrm{u}}}} \to \mathbb{R}^{n_\mathrm{\sche}}$, the so-called scheduling map, such that $\forall (\bar x(t), \bar u(t)) \in \mc{X}\times\mc{U}$:
\begin{align*}
A(\psi(\bar{x}, \bar{u})) &= \tPartial{}{f}{x}(\bar{x}, \bar{u}), \ \ B(\psi(\bar{x}, \bar{u})) = \tPartial{}{f}{u}(\bar{x}, \bar{u}),\\
C(\psi(\bar{x}, \bar{u})) &= \tPartial{}{h}{x}(\bar{x}, \bar{u}), \ \ D(\psi(\bar{x}, \bar{u}))= \tPartial{}{h}{u}(\bar{x}, \bar{u}),
\end{align*}
where $A,\dots,D$ belong to a given function class (affine, polynomial, etc.), implying that $\sche(t) = \psi(\bar x(t),\bar u(t))$, and $\mc{P}\supseteq\psi(\mathcal{X},\mathcal{U})$.
\end{definition}
The convex set $\mc{P}$ is usually a superset of the $\psi$-projected values of possible state and input trajectories (even if $\mathcal{X}$, $\mathcal{U}$ are convex), hence the {D}PV embedding of a nonlinear system introduces conservatism. However, this is considered to be the trade-off for efficiently computable stability and performance analysis of nonlinear systems. To reduce the conservatism of the {D}PV embedding \eqref{eq:sys_LPV} for a given preferred dependency class of $A,B,C,D$ (e.g. affine, polynomial, rational), we can optimize $\psi$ (with minimal $n_\mr{\sche}$) such that $\cohull{\psi(\mc{X},\mc{U})}\setminus\psi(\mc{X},\mc{U})$ has minimal volume \cite{sadeghzadeh2020linear,Toth2010}. Note that the {D}PV embedding serves as an important tool to \emph{convexify} the variation of the matrix inequalities in the analysis. {In turn, that allows to} solve the derived \emph{infinite} set of {LMIs} using a \emph{finite} set of LMIs  {using SDP, e.g., via} polytopic or multiplier based methods \cite{HoffmannWerner2015}.

In case that $\mc{X}\times\mc{U}$ is unbounded, the DPV embedding is often realized on a convex subset $\mf{X}\times\mf{U}$ of $\mc{X}\times\mc{U}$, 
such that there exists a compact and convex $\mc{P}\supseteq\psi(\mf{X},\mf{U})$. In this case, one either requires to add an extra condition of invariance of the system on $\mf{X}\times\mf{U}$ or assume it, which may 
introduce conservatism in the analysis, as not the full behavior of the original primal system is considered. Note that existence of a compact and convex $\mc{P}$, in case of unbounded $\mc{X}\times\mc{U}$, follows when $\tPartial{}{f}{x},\, \tPartial{}{f}{u} ,\, \tPartial{}{h}{x},\,\tPartial{}{h}{u}$ are bounded matrix functions, e.g., if $\tPartial{}{f}{x}=\sin(x)$, with $x\in\mathbb{R}$, we can take $\sche = \psi(x) = \sin(x)\in[-1,1]$.

\shortsection{Example\extver{}{s}}\label{section:example}
This section demonstrates the developed {notions of}  incremental dissipativity theory\footnote{There are some works under review that apply the developed theory in Section \ref{section:mainresults} on a practical example.} {and the analysis tools on} \extver{an example. The extended version \cite{VerhoekKoelewijnToth2020} contains an additional example on the relation between incremental and general dissipativity}{two example systems}.
\begin{example}
Consider a second-order Duffing oscillator given in a state-space form by
\begin{equation}
\left\{
\begin{aligned}
\dot{x}_1(t) & = x_2(t); \\
\dot{x}_2(t) & = -a \,x_2(t) - \left(b + c \, x_1^2(t)\right)x_1(t) + u(t); \\
y (t) & = x_1(t),
\end{aligned}\right.\label{eq:duffingoscillator}
\end{equation}
where $a$ and $b$ represent the linear damping and stiffness, respectively, and $c$ represents the nonlinear stiffness component. The differential form of \eqref{eq:duffingoscillator} is given by
\begin{equation}
\left\{
\begin{aligned}
\delta\dot{x}(t) & = \begin{psmallmatrix} 0 & 1 \\ - b - 3\,c\,x_1^2(t) & -a\end{psmallmatrix}\delta x(t) + \begin{psmallmatrix}0\\1 \end{psmallmatrix}\delta u(t); \\
\delta y (t) & = \begin{psmallmatrix} 1 & 0 \end{psmallmatrix} \delta x(t).
\end{aligned}\right.\label{eq:diffduffingoscillator}
\end{equation}
Moreover, we assume for this system that $(x_1,x_2)\in\ms{X}$, with $\ms{X}=\mc{X}^\mb{R}$, where $\mc{X}=[-\sqrt{2}, \sqrt{2}]\times\mb{R}$, and
\begin{equation*}
u\in\ms{U}:=\ms{L}_2 \cap \setdefinition{\mb{R}^\mb{R}\!}{\text{\eqref{eq:duffingoscillator}
holds and } (x_1,x_2)\in \ms{X}}.
\end{equation*}By choosing $a=3.3$, $b=7.9$, $c=1$, \eqref{eq:duffingoscillator} yields a system with finite \litwo-gain. In this example, we determine the \litwo-gain of the system, using Corollary~\ref{lemma:l2}. Note that the nonlinearity $x_1^2(t)$ in \eqref{eq:diffduffingoscillator} can be captured by using a DPV inclusion $\sche(t) = \psi(x_1(t)) = x_1^2(t)\in[0,2]$. By this substitution, \eqref{eq:L2convex} becomes a matrix inequality linear in $\sche$, which can be reduced to a finite number of LMI constraints at the vertices, due to convexity of $[0,2]$. Solving the resulting SDP (constrained minimization of $\gamma$) yields $M=\begin{psmallmatrix} 0.592 & 0.0896\\0.0896 & 0.0543 \end{psmallmatrix}\posdef 0$ and $\gamma=0.155$. Hence, within less than a second, we know that the nonlinear system is differentially, incrementally and {generally} dissipative {on $\ms{X}$} w.r.t. the supply function \eqref{eq:supply_diff} with $Q=0.155^2$, $R=-1$ and $S=0$, and that it has an \litwo-gain less than 0.155. The system is simulated with two different input signals, given in \eqref{eq:inputs}, for which we know they are in $\ms{U}$.
\begin{subequations}\label{eq:inputs}
\begin{align}
u_1(t) & = 3e^{-0.2 t}\cos\!\left(\pi t\right)\unitstep,\\
u_2(t) & = -2 e^{-0.1 t}\sin\!\left(0.6\pi t+\tfrac{\pi}{4}\right)\unitstep,
\end{align}
\end{subequations}
where $\unitstep$ is the unit step-function. The inputs and the state trajectories are shown in Fig. \ref{fig:dissipative_example1_traj}, which shows that the states stay within the defined state-space $\ms{X}$.
\begin{figure}[!t]
\centering
\includegraphics[width=\linewidth]{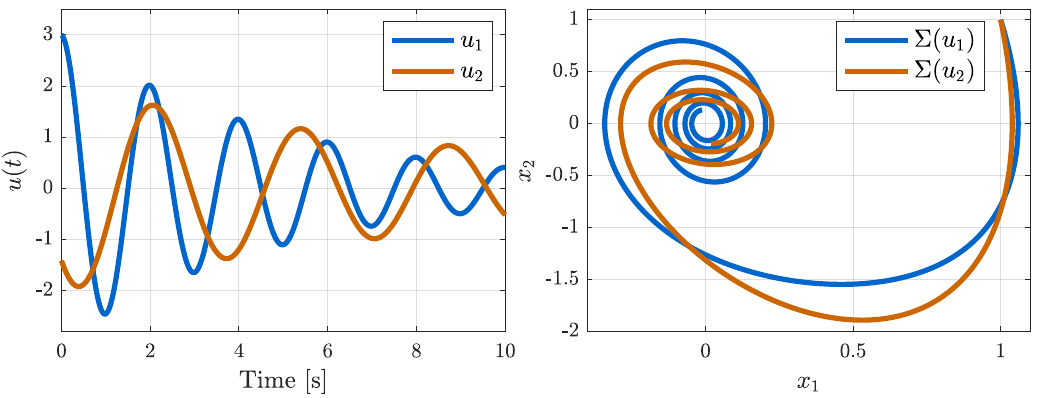}
\spacenafig
\caption{The applied inputs to the system (left) and the resulting state trajectories (right). Both trajectories start at $\big(x_1(0), x_2(0)\big) = (1,1)$.}
\spacenacap
\label{fig:dissipative_example1_traj}
\end{figure}

To verify whether the system is differentially dissipative, considering these specific trajectories, the signals of \eqref{eq:diffduffingoscillator} are substituted in the {DI} for the differential form \eqref{eq:DIE_diff}. The left- and right-hand side of the {DI} \eqref{eq:DIE_diff} are plotted in Fig. \ref{fig:dissipative_example1_diff} corresponding to the system trajectories of Fig. \ref{fig:dissipative_example1_traj}. As can be {seen} in Fig. \ref{fig:dissipative_example1_diff}, the stored energy in the system is always less than the supplied energy plus the initial stored energy, hence the system is differentially dissipative {w.r.t. the considered $\ltwo$-gain supply}.
\begin{figure}[!t]
\centering
\begin{subfigure}[c]{\linewidth}
	\includegraphics[width=\linewidth]{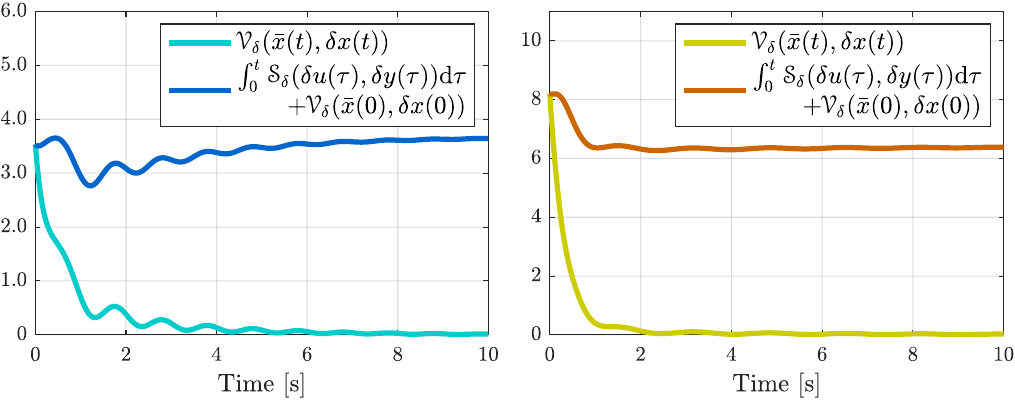}
	\spacenafig
	\caption{Differential dissipativity of the system trajectories with $u_1(t)$ as input (left) and $u_2(t)$ as input (right).}
	\label{fig:dissipative_example1_diff}
\end{subfigure}\\ \vspace{2mm}
\begin{subfigure}[c]{\linewidth}
	\includegraphics[width=\linewidth]{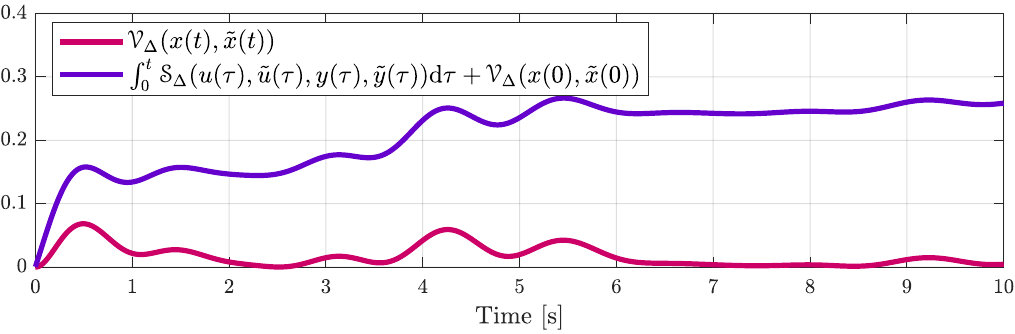}
	\spacenafig
	\caption{Incremental dissipativity based on the system trajectories with $u_1(t)$ and $u_2(t)$ as input.}
	\label{fig:dissipative_example1_inc}
\end{subfigure}\\ \vspace{2mm}
\begin{subfigure}[c]{\linewidth}
	\includegraphics[width=\linewidth]{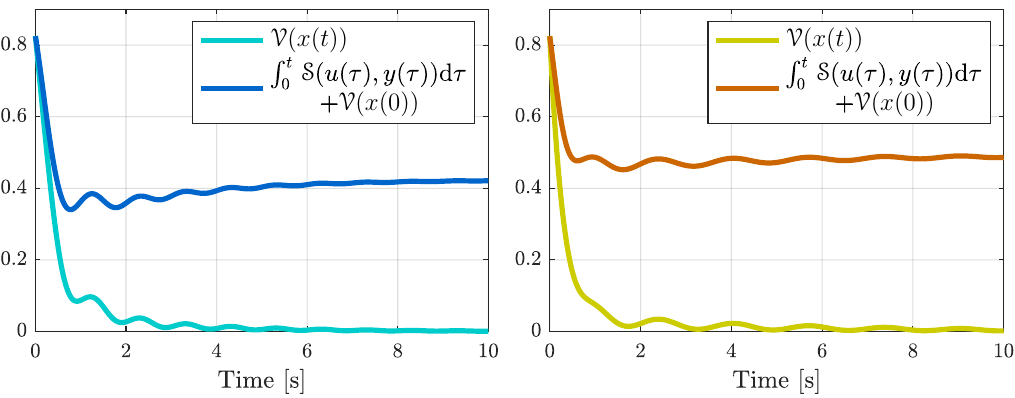}
	\spacenafig
	\caption{General dissipativity of the system trajectories with $u_1(t)$ as input (left) and $u_2(t)$ as input (right).}
	\label{fig:dissipative_example1_gen}
\end{subfigure}
\caption{Simulation results for the different notions of dissipativity for a Duffing oscillator w.r.t. the supply function $\supply(u,y) = \gamma^2 \|u\|^2-\|y\|^2$, corresponding to the (incremental) \ltwo-gain.}
\spacenacap
\label{fig:simresultsexample2}
\end{figure}

Since the system is differentially dissipative 
{it is also} incrementally dissipative. Fig.~\ref{fig:dissipative_example1_inc} shows the incremental dissipation inequality, i.e., the stored energy and the supplied energy between the two trajectories in Fig.~\ref{fig:dissipative_example1_traj}. As can be observed in Fig.~\ref{fig:dissipative_example1_inc}, the stored energy between two trajectories is always less than the supplied energy between two trajectories. Hence, considering these trajectories, the system is incrementally dissipative. Therefore, we can state (based on these two trajectories) that these results correspond to the developed theory. Furthermore, because the supply function is parametrized such that it represents the \litwo-gain of a system, $\gamma=0.155$ is an upper bound for the \litwo-gain of the system \eqref{eq:duffingoscillator}.

Moreover, by Theorem \ref{theorem:incr_normal}, incremental dissipativity implies general dissipativity of the original system \eqref{eq:duffingoscillator}. Fig. \ref{fig:dissipative_example1_gen} {gives} the storage and supply function evolution over time for the two considered trajectories, showing that the original system is dissipative, since the stored energy is always less than the supplied energy. 
\hfill $\blacktriangleleft$ \end{example}
\extver{}{
The next example shows that incremental dissipativity is a \emph{stronger} notion than general dissipativity, if the same type of storage function is considered.
\begin{example}
This example again uses a Duffing oscillator, now with the output equation given by $y(t) = x_2(t)$. With this small modification compared to \eqref{eq:duffingoscillator}, the Duffing oscillator can be written as a port-Hamiltonian system. From \cite{Molero2013}, we take the Hamiltonian function as
\begin{equation}
\mathcal{H}(x) = \tfrac{1}{2}x_2^2 + \tfrac{1}{2}b \, x_1^2 + \tfrac{1}{4} c \, x_1^4.
\end{equation}
The resulting port-Hamiltonian form of this system is
\begin{equation*}
\left\{
\begin{aligned}
\dot{x}(t) & = \underbrace{\left[\begin{pmatrix} 0 & 1 \\ -1 & 0 \end{pmatrix} - \begin{pmatrix} 0 & 0 \\ 0 & a \end{pmatrix} \right]}_{[J(x)-R(x)]} \underbrace{\begin{pmatrix} \Partial{}{\mc{H}}{x_1} \\ \Partial{}{\mc{H}}{x_2} \end{pmatrix}}_{\nabla \mc{H}(x)} + \underbrace{\begin{pmatrix} 0 \\ 1 \end{pmatrix}}_{B(x)}u(t); \\
y (t) & = B(x)^\top\nabla \mc{H}(x).
\end{aligned}\right.
\end{equation*}
Since a port-Hamiltonian system is always dissipative w.r.t. the supply function $\supply(u(t),y(t)) = 2u^{\!\top}\!(t) y(t)$, we know that the dissipation inequality holds for all trajectories. 
\begin{figure}[!t]
\centering
\begin{subfigure}[c]{\linewidth}
	\includegraphics[width=\linewidth]{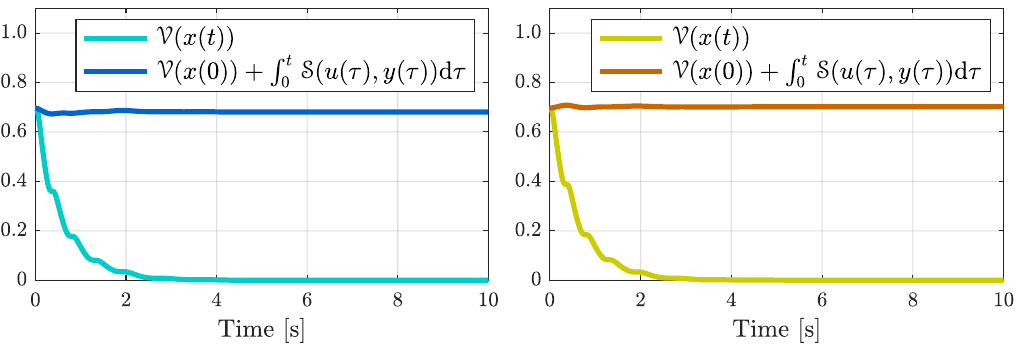}
	\caption{General dissipativity of the first (left) and second (right) trajectory, here the y-axis is normalized unitarily.}
	\label{fig:dissipative_example2_gen}
\end{subfigure}\\ \vspace{1mm}
\begin{subfigure}[c]{\linewidth}
	\includegraphics[width=\linewidth]{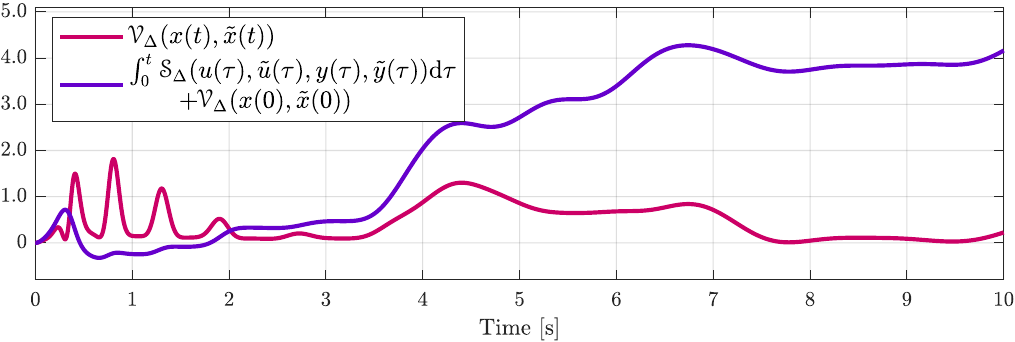}
	\caption{Incremental dissipativity.\hfill}
	\label{fig:dissipative_example2_inc}
\end{subfigure}
\caption{Simulation results for a Duffing oscillator which is passive, but not incrementally passive, when the same storage function (the Hamiltonian function) is used, i.e., $\mc{H}(x)$ for general passivity and $\mc{H}(x-\tilde x)$ for incremental passivity.}
\label{fig:dissipation_example2}
\end{figure}
Moreover, this supply function indicates passivity, hence the port-Hamiltonian system is passive. By choosing $a=1.3, \, b=7.9, \, c=3$, yields a system that is passive, but not incrementally passive, when the same Hamiltonian is used, i.e. $\mc{H}(x-\tilde x)$. The two plots in Fig. \ref{fig:dissipative_example2_gen} show the (normalized) dissipation inequality for two arbitrary inputs, and indeed the energy in the system is less than the supplied energy to the system. Hence, the system is passive. However, when incremental dissipativity is examined by subtracting both trajectories, the plot in Fig. \ref{fig:dissipative_example2_inc} is obtained. For some time-interval, the energy in the system is more than the energy supplied to the system, hence the system is not incrementally passive w.r.t. the supply function $\supply(u,\tilde{u},y,\tilde{y}) = 2\left(u-\tilde{u}\right)^\top \left(y-\tilde{y}\right)$ and storage function $\mc{H}(x-\tilde x)$. This shows that incremental dissipativity is a stronger notion than general dissipativity, when the storage function has the same complexity. Note that the system might be incrementally dissipative for some different storage function.
\hfill $\blacktriangleleft$ \end{example}}
\shortsection{Conclusions}\label{section:conclusions} 
\extver{}{\vspace{-1mm}}
In this paper, we established the link between general dissipation theory, incremental dissipativity analysis and differential dissipativity analysis for nonlinear systems. Moreover, {we have given results on general quadratic incremental performance notions and}  parameter-varying inclusion based computation tools to analyze the different notions of dissipativity in a convex setting by SDPs. The established link gives us a generic framework to analyze stability and performance of a nonlinear system from a global perspective. Finally, the presented computation tools allow to efficiently analyze global stability and performance of a rather general class of nonlinear systems. {These} results open up the possibility to establish controller synthesis based on PV inclusions of the differential form, such that we can synthesize (nonlinear) controllers for nonlinear systems with incremental stability  and performance guarantees of the closed-loop behavior. For future work, we aim to extend the developed theory for discrete-time and time-varying nonlinear systems. 
\extver{\vspace{-2mm}}{}
\begin{ack}
\extver{\vspace{-2mm}}{}
The authors would like to thank prof. dr. A. J. van der Schaft for commenting on the draft version of this paper.
\end{ack}

\appendix
\shortsection{Norm Integral Inequality}\label{sec:appendix}
From \cite[Lem. 16]{koelewijn2021incremental},
we have the following useful lemma.
\begin{lem}\label{lem:appendix}
	For $0\negdef M = M^\top \in\mathbb{R}^{n\times n}$ and a continuous function $\phi:[0,1]\rightarrow\mathbb{R}^n$, it holds that
\begin{equation}\label{eq:lem:appendix}
 \left({\textstyle\int_0^1}\phi(t)\dif t\right)^{\!\top} \!\! M \left({\textstyle\int_0^1}\phi(t)\dif t\right)\leq \integral{0}{1}\extver{}{\!} \phi^{\hspace{-0.2mm}\top}\hspace{-0.8mm}(t) M\,\phi(t)\,\dif t.\hspace{-1mm}
\end{equation} \end{lem}

\extver{}{
\section{Proofs}
\shortsubsection{Detailed proof of Corollary \ref{lemma:l2}}\label{appendix:l2}
\begin{proof}
First, we show that if for all $(\bar x, \bar u)\in\pi_{x,u}\mc{B}_2$,  \eqref{eq:L2convex} holds with $M\posdef0$, the system \eqref{eq:sys} is incrementally dissipative w.r.t. the supply function
\begin{equation}
\supply_\Delta(u,\tilde{u}, y, \tilde{y}) = \gamma^2 \norm{u-\tilde{u}}^2 - \norm{y - \tilde{y}}^2, \label{eq:L2supplyfunction}
\end{equation}
and storage function
\begin{equation}\label{eq:L2storage}
\storage_\Delta (x,\tilde{x}) = (x-\tilde{x})^\top M (x-\tilde{x}), \quad M\posdef0.
\end{equation}
Applying the Schur-complement on \eqref{eq:L2convex} with $M\posdef0$ gives
\begin{multline}\label{eq:L2expression}
M\posdef 0, \quad (*)^\top  \begin{pmatrix} 0  &  M\\M  &  0 \end{pmatrix}   \begin{pmatrix} 
I   &   0 \\ 
A \left(\bar{x},\bar{u}\right)   &   B \left(\bar{x},\bar{u}\right)
\end{pmatrix} - \\
(*)^\top   \begin{pmatrix} \gamma^2I  &  0\\0  &  -I \end{pmatrix} \begin{pmatrix} 
0    &    I \\ 
C \left(\bar{x},\bar{u}\right)    &     D \left(\bar{x},\bar{u}\right)
\end{pmatrix}   \preccurlyeq 0.
\end{multline}
From this we can get the $QSR$-triple $(\gamma^2 I, 0, -I)$, i.e., $R\negdef0$. Hence, from Corollary \ref{coro:incrementaldissip} we have that the system \eqref{eq:sys} is incrementally dissipative w.r.t. the storage function \eqref{eq:L2storage}, and supply function
\begin{equation}\label{eq:L2supplyfunction2}
\supply_\Delta(u,\tilde{u}, y, \tilde{y}) = \diffstate{u}{y}\shrtranspose\!\!\begin{pmatrix}\gamma^2I&0\\0&-I\end{pmatrix}\diffstate{u}{y},
\end{equation}
which is equivalent to the supply function \eqref{eq:L2supplyfunction}. Therefore, if \eqref{eq:L2convex} holds with $M\posdef0$, the system \eqref{eq:sys} is incrementally dissipative w.r.t. supply function \eqref{eq:L2supplyfunction}. Among the lines of \cite[Prop. 3.1.7]{L2PassiveSchaft2017}, we now show that this implies that $\|\Sigma\|_{\litwo}\!\le\gamma$.

Note that if the system is incrementally dissipative w.r.t. \eqref{eq:L2supplyfunction}, there exists a ${\bar{\storage}_\Delta}(t) :={\storage_\Delta}(x(t),\tilde{x}(t))\ge 0$ such that for all $t\ge 0$, $(x(0)-\tilde{x}(0))$ and $(u(\cdot)-\tilde{u}(\cdot))$ we can write
\begin{multline}
-{\bar{\storage}_\Delta}(0)\le {\bar{\storage}_\Delta}(t)-{\bar{\storage}_\Delta}(0) \le\\ \int_{0}^{t}\!\! \left(\gamma^2 \norm{u(\tau)-\tilde{u}(\tau)}^2 - \norm{y(\tau) - \tilde{y}(\tau)}^2\right)\dif \tau,
\end{multline}
which is equivalent to
\begin{equation}
\int_{0}^{t}\!\!\norm{y(\tau)-\tilde{y}(\tau)}^2 \dif \tau \le \gamma^2\!\! \int_{0}^{t}\!\! \norm{u(\tau)-\tilde{u}(\tau)}^2 \dif \tau + {\bar{\storage}_\Delta}(0).\label{eq:l2expressioninproof}
\end{equation}
Note that by Definition \ref{DefinitionL2}, $x(0)=\tilde{x}(0)$, i.e., $x(0)-\tilde{x}(0)=0$. Therefore, from \eqref{eq:L2storage} we have that ${\bar{\storage}_\Delta}(0)=0$. When taking $t\to\infty$ in  \eqref{eq:l2expressioninproof}, an inequality in the signal norms is obtained,
\begin{equation}
\norm{y - \tilde{y}}^2_2 \le \gamma^2 \norm{u-\tilde{u}}^2_2 \ \Leftrightarrow \ \frac{\norm{y - \tilde{y}}^2_2}{\norm{u-\tilde{u}}^2_2}\le \gamma^2.
\end{equation}
Taking the square root on both sides and taking the supremum over all $0<\norm{u-\tilde{u}}_2<\infty$, yields that $\|\Sigma\|_{\litwo}\!\le\gamma$, proving the statement.
\end{proof}
%
%
\shortsubsection{Detailed proof of Corollary \ref{corollary:l1}}\label{appendix:l1}
\begin{proof}
The concept of this proof is based on \cite[Section~10.3]{SchererL1norm} and \cite[Section~3.3.5]{LMICourse}. Pre- and post-multiplying \eqref{eq:L1LMI1} with $(\begin{matrix} \delta x^\top &  \delta u^\top \end{matrix})$ and $(\begin{matrix} \delta x^\top &  \delta u^\top \end{matrix})^\top$, respectively, yields that for $(\delta x, \delta u)\neq0$,
\begin{equation}\label{eq:pfs:col17:1}
	\delta \dot x^\top M \delta x + \delta x^\top M \delta \dot x + \kappa \delta x^\top M \delta x - \mu \delta u^\top\delta u < 0,
\end{equation}
omitting dependence on $t,\lambda$. For 
\[{\bar{\storage}_\delta}(\delta x(t,\lambda)):=\delta x(t,\lambda)^\top M \delta x(t,\lambda),\]
\eqref{eq:pfs:col17:1} is equivalent to
\begin{equation}\label{eq:pf:linf:1}
\hspace{-0.2mm}\differ{}{t}\Big({\bar{\storage}_\delta}\big(\delta x(t,\lambda)\big)\!\Big) + \kappa {\bar{\storage}_\delta}(\delta x(t,\lambda)) \unaryminus \mu \norm{\delta u(t,\lambda)}^2\!<\! 0.\hspace{-1mm}
\end{equation}
Furthermore, note that from Definition \ref{def:incrLinf} we have
\begin{multline}
	\delta x (0,\lambda) = \Partial{}{\bar{x}_0(\lambda)}{\lambda} = \tPartial{}{}{\lambda}(\tilde{x}_0+\lambda(x_0-\tilde{x}_0)) = \\ = (x_0-\tilde{x}_0) = 0 \ \implies \ {\bar{\storage}_\delta}(\delta x(0,\lambda)) = 0.\label{eq:pfs:linf:12}
\end{multline}
Together with \eqref{eq:pf:linf:1}, \eqref{eq:pfs:linf:12}, Gr\"onwall's Lemma \cite[Lemma A.1]{KhalilNL} implies that
\begin{equation}\label{eq:pf:linf:2}
{\bar{\storage}_\delta}(\delta x(t,\lambda))<\mu \! \int_{0}^t e^{-\kappa(t-\tau)}\norm{\delta u(\tau,\lambda)}^2 \dif \tau.
\end{equation}
Moreover, we can define an upper bound for the integral in \eqref{eq:pf:linf:2} as follows, 
\begin{multline}
\mu \! \int_{0}^t e^{-\kappa(t-\tau)}\norm{\delta u(\tau,\lambda)}^2 \dif \tau, \\
\le \,  \mu \norm{\delta u(\lambda)}_\infty^2 \int_{0}^t e^{-\kappa(t-\tau)} \dif \tau,\\
\le \,  \frac{\mu}{\kappa} \norm{\delta u(\lambda)}_\infty^2,
\end{multline}
and therefore $\forall t\ge 0$: $\kappa {\bar{\storage}_\delta}(\delta x(t,\lambda)) < \mu \norm{\delta u(\lambda)}_\infty^2$. As $\kappa {\bar{\storage}_\delta}(\delta x(t,\lambda))\ge 0$, we know that the following is true,
\begin{multline}\label{eq:pf:linf:3}
	\int_0^1\!\! \kappa {\bar{\storage}_\delta}(\delta x(t,\lambda))\dif\lambda = \kappa\! \int_0^1\!\!  \delta x(t,\lambda)^\top M \delta x(t,\lambda)\dif\lambda < \\ < \int_0^1 \!\!\mu \norm{\delta u(\lambda)}_\infty^2 \dif\lambda = \mu \norm{ u-\tilde{u}}_\infty^2.
\end{multline}
The latter equality comes from the fact that $\delta u(\lambda) = \tPartial{}{\bar{u}}{\lambda}=u-\tilde{u}$. We will use the bound in \eqref{eq:pf:linf:3} later in the proof.

Consider now the second inequality \eqref{eq:L1LMI2}, which can be rewritten using the Schur complement as 
\begin{equation*}
\frac{1}{\gamma} \begin{pmatrix} * \end{pmatrix}\shrtranspose\begin{pmatrix}
C \left(\bar{x},\bar{u}\right) & D \left(\bar{x},\bar{u}\right)
\end{pmatrix} \prec \begin{pmatrix}
\kappa M & 0 \\ 0 & (\gamma - \mu) I 
\end{pmatrix} .
\end{equation*}
Pre- and post-multiplying the latter with $(\begin{matrix} \delta x^\top &  \delta u^\top \end{matrix})$ and $(\begin{matrix} \delta x^\top &  \delta u^\top \end{matrix})^\top$, respectively, gives that,
\begin{multline}\label{eq:pf17:1}
\tfrac{1}{\gamma} \,\delta y(t,\lambda)^\top \delta y(t,\lambda) \le \kappa\, \delta x(t,\lambda)^\top M\delta x(t,\lambda) \\ +(\gamma-\mu)\delta u(t,\lambda)^\top\delta u(t,\lambda).
\end{multline}
As \eqref{eq:pf17:1} only consists of positive terms, we know the following is true as well,
\begin{multline}\label{eq:pf17:2}
\!\!\int_0^1\!\!\!\Big(\tfrac{1}{\gamma} \,\delta y(t,\lambda)^\top \delta y(t,\lambda)\Big)\dif\lambda \le \!\!\int_0^1\!\!\!\Big( \kappa\, \delta x(t,\lambda)^\top M\delta x(t,\lambda) \\ +(\gamma-\mu)\delta u(t,\lambda)^\top\delta u(t,\lambda)^\top\Big)\dif\lambda.
\end{multline}
Using Lemma \ref{lem:appendix}, we know that the left-hand side of the inequality in \eqref{eq:pf17:2}, is bounded from below by
\[ \tfrac{1}{\gamma}\left({\textstyle\int_0^1}\delta y(t,\lambda)\dif \lambda\right)^\top \!\!\left({\textstyle\int_0^1}\delta y(t,\lambda)\dif \lambda\right)= \tfrac{1}{\gamma}(*)^\top(y(t)-\tilde{y}(t)).\]
Again due to the selection of the path parametrization of $\bar u$, we have that the last term in \eqref{eq:pf17:2} is not dependent on $\lambda$, as $\delta u(t,\lambda) = u(t)-\tilde{u}(t)$. Hence, we have that \eqref{eq:pf17:2} can be rewritten to
\begin{multline}\label{eq:pf17:3}
\tfrac{1}{\gamma}\norm{y(t)-\tilde{y}(t)}^2 \le \kappa\!\!\int_0^1\!\!\!\delta x(t,\lambda)^\top M\delta x(t,\lambda)\dif\lambda \\ +(\gamma-\mu)\norm{u(t)-\tilde{u}(t)}^2.
\end{multline}
By substituting the inequality \eqref{eq:pf:linf:3} in the inequality \eqref{eq:pf17:3}, we obtain the following relationship,
\begin{equation*}
\tfrac{1}{\gamma} \norm{y(t)-\tilde{y}(t)}^2 < \mu \norm{u-\tilde{u}}_\infty^2+(\gamma-\mu)\norm{u(t)-\tilde{u}(t)}^2.
\end{equation*}
By taking the supremum over all $t\ge 0$, we infer
\begin{equation*}
\norm{y-\tilde{y}}_\infty^2 < \gamma^2 \norm{u-\tilde{u}}_\infty^2.
\end{equation*}
Taking the square root on both sides and dividing both sides by $\norm{u-\tilde{u}}_\infty$ yields
\begin{align*}
\frac{\norm{y-\tilde{y}}_\infty}{\norm{u-\tilde{u}}_\infty}< \gamma.
\end{align*}
Taking the supremum over $0<\norm{u-\tilde{u}}_\infty<\infty$ yields the definition of the incremental \linf-gain, proving the claim. 
\end{proof}
%
%
\shortsubsection{Detailed proof of Corollary \ref{corollary:h2}}\label{appendix:h2}
\begin{proof}
The concept of this proof is based on \cite[Section~3.3.4]{LMICourse}. We have that the first matrix inequality \eqref{eq:H2LMI1} is equivalent to the condition for differential dissipativity of a system with the supply function
\[\supply_\delta(\delta u(t,\lambda),\delta y(t,\lambda)) = \gamma\norm{\delta u(t,\lambda)}^2=\gamma\norm{\tPartial{}{\bar{u}(t,\lambda)}{\lambda}}^2,\]
as a result from Theorem~\ref{theorem:differential}. From Definition \ref{def:h2} we have that $x(0)-\tilde{x}(0)=0$. Considering $\delta x(0,\lambda) = \tPartial{}{\bar{x}_0(\lambda)}{\lambda}= \tPartial{}{}{\lambda}(\tilde{x}(0)+\lambda(x(0)-\tilde{x}(0)))=x(0)-\tilde{x}(0)$, we can conclude that $\delta x(0,\lambda)=0$. Therefore, it holds for all $t\ge 0$ that
\begin{equation*}
\underbrace{\delta x(t,\lambda)\shrtranspose M \delta x(t,\lambda)}_{={\bar{\storage}_\delta}(\delta x(t,\lambda))} \le \gamma \! \int_{0}^{t}\!\norm{u(\tau)-\tilde{u}(\tau)}^2\dif \tau,
\end{equation*}
using the fact that $\tPartial{}{\bar{u}(t,\lambda)}{\lambda}= \tPartial{}{}{\lambda}(\tilde{u}+\lambda(u(t)-\tilde{u}(t)))=u(t)-\tilde{u}(t)$. Moreover, as ${\bar{\storage}_\delta}(\delta x(t,\lambda))\ge0$, we have that the following holds true as well
\begin{equation*}
\int_0^1{\bar{\storage}_\delta}(\delta x(t,\lambda))\,\dif\lambda \le \int_0^1 \left(\gamma \! \int_{0}^{t}\!\!\left\|u(\tau)-\tilde{u}(\tau)\right\|^2\dif \tau\right)\dif\lambda,
\end{equation*}
that is,
\begin{equation}\label{eq:pf:h2:1}
\int_0^1{\bar{\storage}_\delta}(\delta x(t,\lambda))\,\dif\lambda \le \gamma \! \int_{0}^{t}\!\!\left\|u(\tau)-\tilde{u}(\tau)\right\|^2\dif \tau.
\end{equation}

Consider now the second matrix inequality \eqref{eq:H2LMI2}, which can be rewritten using the Schur complement, such that for $\gamma > 0$ it holds that
\begin{equation}\label{eq:pfs:h2:1}
	M - \tfrac{1}{\gamma}C \left(\bar{x},\bar{u}\right)\shrtranspose C \left(\bar{x},\bar{u}\right) \succ  0.
\end{equation}
The matrix inequality \eqref{eq:pfs:h2:1} is equivalent to stating that for some $\varepsilon>0$,
\begin{equation}\label{eq:pfs:h2:2}
	C \left(\bar{x},\bar{u}\right)\shrtranspose C \left(\bar{x},\bar{u}\right) \negsemidef \left(\gamma-\varepsilon\right)M.
\end{equation}
Pre- and post-multiplication of \eqref{eq:pfs:h2:2} with $\delta x(t,\lambda)^\top$ and $\delta x(t,\lambda)$, respectively, gives
\begin{equation}\label{eq:pf:h2:2}
	\delta y(t,\lambda)^\top \delta y(t,\lambda) \le (\gamma-\varepsilon){\bar{\storage}_\delta}(\delta x(t,\lambda)),
\end{equation}
as $\tPartial{}{\bar{y}(t,\lambda)}{\bar{u}}=\tPartial{}{h(\bar{x},\bar{u})}{\bar{u}}=0$ for all $\bar{x}(t,\lambda),\bar{u}(t,\lambda)\in\pi_{x,u}\mc{B}$. Clearly, if \eqref{eq:pf:h2:2} holds, we know that
\begin{equation}\label{eq:pf:h2:3}
	\int_0^1\delta y(t,\lambda)^\top \delta y(t,\lambda)\,\dif\lambda \le \int_0^1(\gamma-\varepsilon){\bar{\storage}_\delta}(\delta x(t,\lambda))\,\dif\lambda,
\end{equation}
holds. As in the proof for the \liinf-gain, we use Lemma \ref{lem:appendix} to obtain a lower bound for the left-hand side of \eqref{eq:pf:h2:3}. The lower bound is
\begin{multline*}
	\left(\int_0^1\delta y(t,\lambda)\dif\lambda\right)^\top\left(\int_0^1\delta y(t,\lambda)\dif\lambda\right) = \\ (y(t)-\tilde{y}(t))^\top (y(t)-\tilde{y}(t)) = \norm{y(t)-\tilde{y}(t)}^2,
\end{multline*}
and gives that \eqref{eq:pf:h2:3} implies that
\begin{equation}\label{eq:pf:h2:4}
	\norm{y(t)-\tilde{y}(t)}^2 \le (\gamma-\varepsilon)\int_0^1{\bar{\storage}_\delta}(\delta x(t,\lambda))\,\dif\lambda.
\end{equation}
Substituting \eqref{eq:pf:h2:1} in \eqref{eq:pf:h2:4} gives that for all $t\ge 0$
\begin{multline}
\norm{y(t)-\tilde{y}(t)}^2 \le (\gamma-\varepsilon)\int_0^1{\bar{\storage}_\delta}(\delta x(t,\lambda))\,\dif\lambda \\
 \le (\gamma-\varepsilon)\gamma \! \int_{0}^{t}\!\!\left\|u(\tau)-\tilde{u}(\tau)\right\|^2\dif \tau \qquad \\
  \quad \le (\gamma^2-\varepsilon\gamma) \! \int_{0}^{t}\!\!\left\|u(\tau)-\tilde{u}(\tau)\right\|^2\dif \tau \\ 
 < \gamma^2 \! \int_{0}^{\infty}\!\!\!\left\|u(\tau)-\tilde{u}(\tau)\right\|^2\dif \tau.
\end{multline}
Taking the supremum over $t\ge 0$ gives 
\begin{equation}\label{eq:pfs:h2:nrm}
\norm{y-\tilde{y}}_\infty^2 < \gamma^2 \norm{u-\tilde{u}}_2^2.
\end{equation}
By taking the square root, and division by $\norm{u-\tilde{u}}_2$ of the expression \eqref{eq:pfs:h2:nrm} yields
\begin{equation*}
\frac{\norm{y-\tilde{y}}_\infty}{\norm{u-\tilde{u}}_2\hfill} < \gamma.
\end{equation*}
Taking the supremum over $0<\norm{u-\tilde{u}}_2<\infty$ yields the definition of the generalized incremental \htwo-norm, proving the statement.
\end{proof}
}

\bibliographystyle{plain}
\bibliography{autoref}

\vspace*{1em}
\extver{\renewcommand{\baselinestretch}{0.83}}{}
\selectfont

\biographytwo{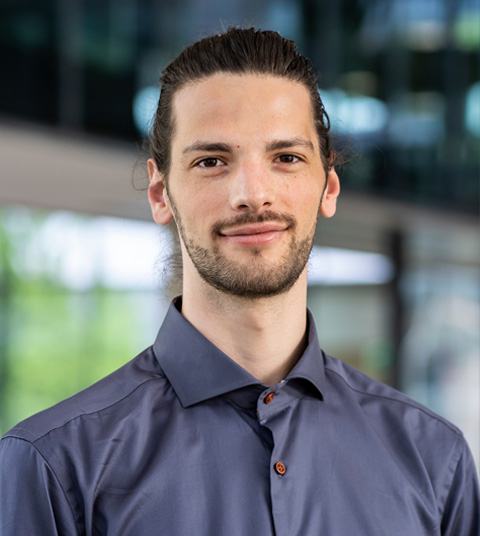}{Chris Verhoek}{received his BSc degree in Mechatronics from the Avans University of Applied Sciences and MSc degree (Cum Laude) in Systems and Control from the Eindhoven University of Technology, in 2017 and 2020 respectively. His MSc thesis was selected as best thesis of the Electrical Engineering department in the year 2020. 
He is currently pursuing a Ph.D. degree at the Control Systems Group, Dept. of Electrical Engineering, Eindhoven University of Technology. His main research interests include (data-driven) analysis and control of nonlinear and LPV systems and learning-for-control techniques with stability and performance guarantees.}
%
%
%
\biographytwo{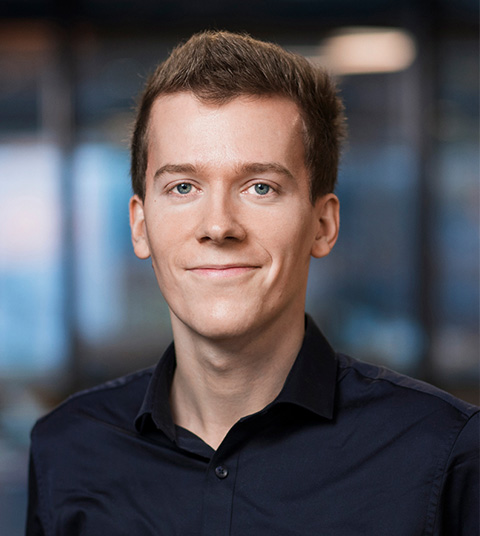}{Patrick J. W. Koelewijn}{received his BSc degree in Automotive and MSc degree in Systems and Control from the Eindhoven University of Technology, both Cum Laude, in 2016 and 2018 respectively. 
During his Master's degree he spent three months at the Institute of Control Systems at the Hamburg University of Technology (TUHH). 
He is currently pursuing a Ph.D. degree at the Control Systems Group, Department of Electrical Engineering, Eindhoven University of Technology. His main research interests include analysis and control of nonlinear and LPV systems, optimal and nonlinear control, and machine learning techniques.}
\biographytwo{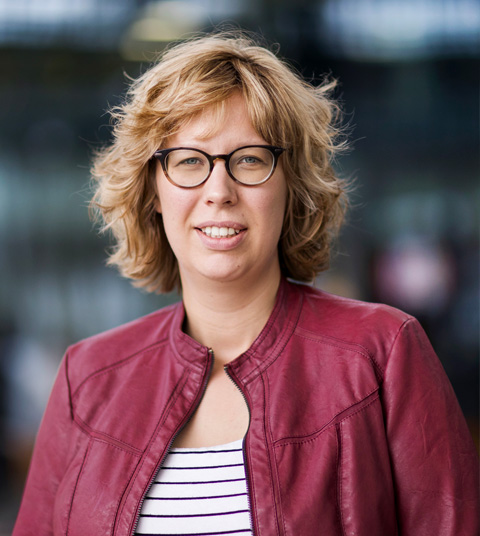}{Sofie Haesaert}{ received the
B.Sc. degree cum laude in mechanical engineering and the M.Sc. degree cum laude in
systems and control from the Delft University of Technology, Delft, The Netherlands, in
2010 and 2012, respectively, and the Ph.D.
degree from Eindhoven University of Technology (TU/e), Eindhoven, The Netherlands, in
2017.
She is currently an Assistant Professor with
the Control Systems Group, Department of
Electrical Engineering, TU/e. From 2017 to 2018, she was a Postdoctoral Scholar with Caltech. Her research interests are in the identification, verification, and control of cyber-physical systems for temporal logic
specifications and performance objectives.}
\biographytwo{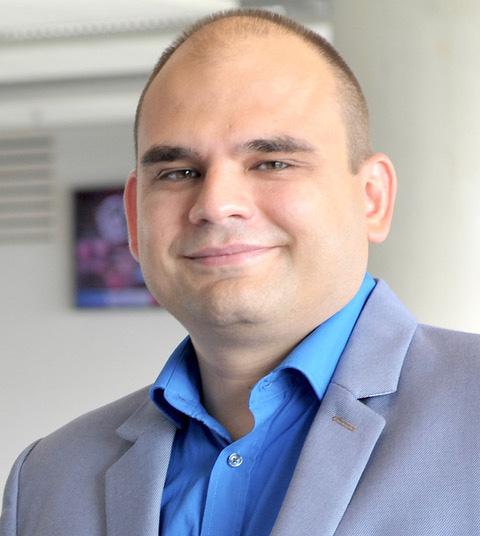}{Roland T\'oth}{received his Ph.D. degree with cum laude distinction at the Delft Center for Systems and Control (DCSC), Delft University of Technology (TUDelft), Delft, The Netherlands in 2008.  He was a Post-Doctoral Research Fellow at TUDelft in 2009 and Berkeley in 2010. He held a position at DCSC, TUDelft in 2011-12. Currently, he is an Associate Professor at the Control Systems Group, Eindhoven University of Technology and a Senior Researcher at SZTAKI, Budapest, Hungary. His research interests are in identification and control of linear parameter-varying (LPV) and nonlinear systems, developing machine learning methods with performance and stability guarantees for modelling and control, model predictive control and behavioral system theory. }

\end{document}